\documentclass{ws-ijbc}
\usepackage{ws-rotating} 
\usepackage{graphicx}
\usepackage{epstopdf}
\begin{document}

\catchline{}{}{}{}{} 

\markboth{X. Tang and S. Mandal}{Digital Communication using Synchronized Hyperchaotic Maps}

\title{Digital Communication using Synchronized Hyperchaotic Maps}

\author{Xinyao Tang and Soumyajit Mandal}
\address{Department of Electrical Engineering and Computer Science,\\ Case Western Reserve University,\\ 
10900 Euclid Avenue, Cleveland, OH 44106, USA\\
sxm833@case.edu}
%

\maketitle

\begin{history}
\received{(to be inserted by publisher)}
\end{history}

\begin{abstract}
This paper describes the analysis and practical implementation of synchronized hyperchaotic maps for private communication of digital data. The data is transmitted using chaotic masking and demodulated using a matched filter (integrate and dump) receiver, which is shown to be nearly optimal in this case. Simulation results were validated by implementing two maps on circuit boards using high-speed discrete components. Experimental results show a bit error rate (BER) of $\approx 2\times 10^{-6}$ at a bit rate of 10~kbps and a clock frequency of 0.5~MHz, which is sufficient for high-fidelity real-time speech and image transmission without additional error control coding.
\end{abstract}


\section{Introduction}
\noindent Hyperchaotic systems exhibit chaotic attractors with more than one positive Lyapunov exponent~\cite{Rossler1979a}. Their outputs can be designed to have short correlation times, high entropy (i.e., low predictability), flat (i.e., white) spectra, and fast synchronization rates, making them attractive as masking signals for private and secure communications~\cite{Grassi1999,Hassan2014,Wang2015}. General schemes for generating hyperchaos via state-feedback control have been developed for both discrete- \cite{Chen2003,Chen2006} and continuous-time \cite{Shen2014,Shen2014b} systems. Since the first experimental observation of hyperchaos in an analog electronic circuit in 1986 ~\cite{Matsumoto1986}, several other implementations have been described ~\cite{Tsubone1997,Carroll1998,Elwakil1999,Grassi2002,Shen2014,Volos2017}. Most of these have emulated continuous-time systems by using low-speed op-amps, resulting in a useful bandwidth of several kHz. Fully-digital implementations based on field-programmable gate arrays (FPGAs) have also been reported~\cite{Chen2011}.

Synchronized hyperchaotic systems are desirable for communications since they enable a variety of synchronous schemes that have better performance than asynchronous schemes such as differential chaos shift keying (DCSK)~\cite{Kolumban1996}. Fortunately, both continuous- and discrete-time hyperchaotic systems can be synchronized by a scalar transmitted signal~\cite{Peng1996,Pecora1997}. Such synchronized systems have been used for communicating analog signals~\cite{Pecora1997}. In particular, the information signal can be ``masked'' with one of the dynamical variables in the transmitter using either linear or nonlinear operations: linear masking operations (e.g., addition) are simple to implement, while nonlinear operations (e.g., XOR) provide more security. In either case, the masked information can be recovered at the receiver using the method of \emph{synchronous substitution}, which is equivalent to standard linear feedback control~\cite{Carroll1996}. Observer-based methods can also be used~\cite{Grassi2002,Grassi2012,Filali2014}. This paper builds on such prior work by focusing on communicating digital signals (bit streams) using synchronized hyperchaotic maps.

This paper is organized as follows. Section~\ref{sec:analysis} describes and analyzes the proposed hyperchaotic system and synchronization scheme. A circuit implementation of the system is discussed in  Section~\ref{sec:design}. Simulation and experimental results on synchronization and digital communication are described in Sections~\ref{sec:sim} and \ref{sec:expt}, respectively. Finally, Section~\ref{sec:conclusion} concludes the paper.

\section{Analysis}
\label{sec:analysis}
\begin{figure}[h]
\begin{center}
\includegraphics[width=0.48\textwidth]{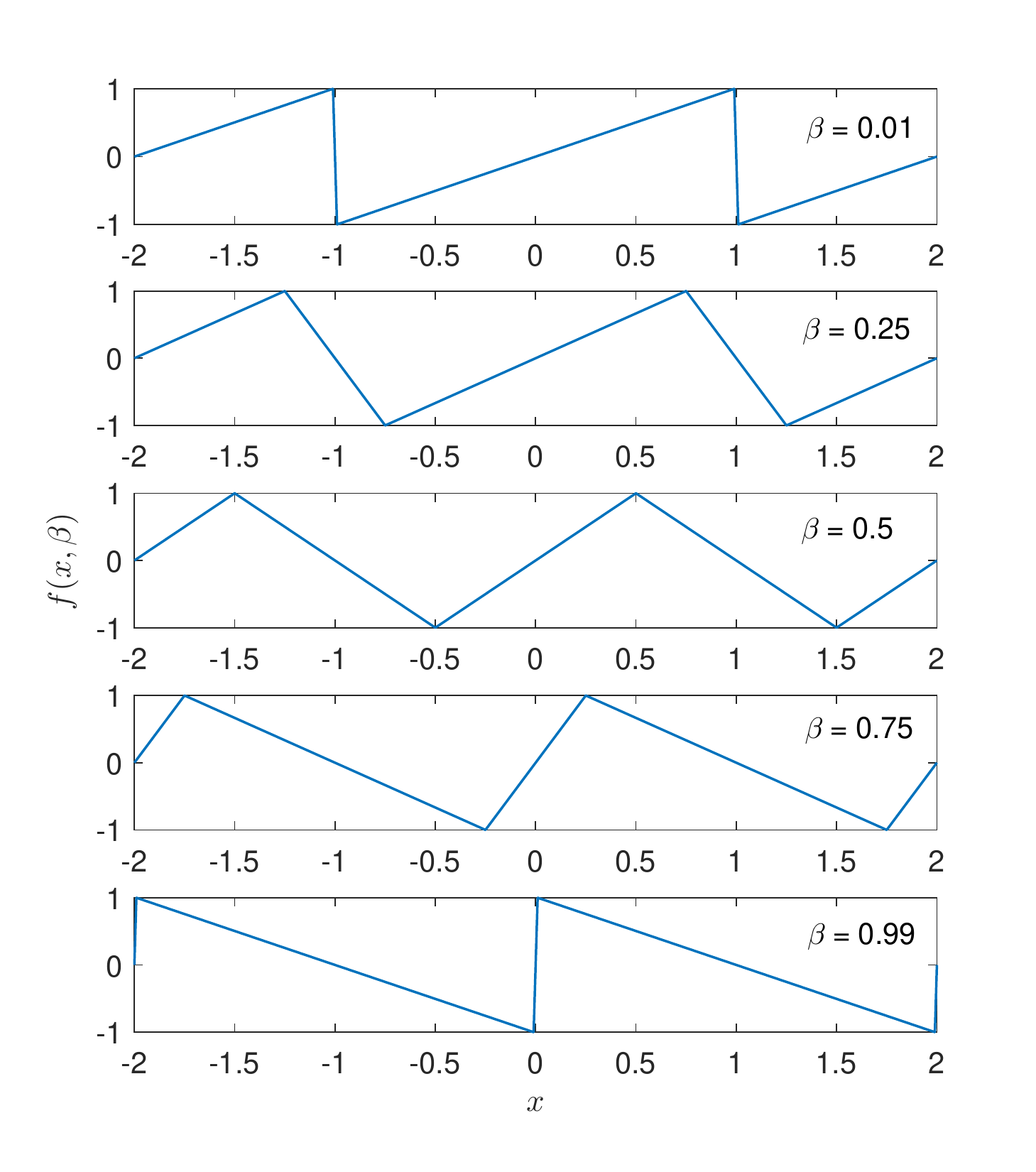}
\end{center}
\caption{The proposed folding function $f(x,\beta)$ for several values of the asymmetry parameter $\beta$.}
\label{fig:hyperchaos_map_2_plots}
\end{figure}

Here we consider a class of discrete-time hyperchaotic systems that include i) a linear transformation that locally expands the volume of phase space along several dimensions (thus resulting in multiple positive Lyapunov exponents); and ii) a nonlinear map that ``folds'' trajectories back into a confined region of phase space whenever they leave. These systems are defined as
\begin{equation}
\mathbf{x}_{n+1}=f(\mathbf{A}\mathbf{x}_{n},\beta),
\label{system_eqn}
\end{equation}

\noindent where $\mathbf{A}$ is the linear transformation matrix, $f(x,\beta)$ is the folding function, and $0\leq \beta\leq 1$ a constant. A variety of folding functions can be used; here we consider $f(x,\beta)$ to be the generalized tent map, which is defined as
\begin{equation}
\begin{array}{*{20}{l}}
{f(x,\beta ) = \left\{ {\begin{array}{*{20}{l}}
{g(x)/(1 - \beta ),}&{{\rm{if}}\;\left| {g(x)} \right| \le (1 - \beta )\;{\rm{and}}\;\beta  < 1}\\
{\left( {1 - g(x)} \right)/\beta ,}&{{\rm{if}}\;\;g(x) > (1 - \beta )\;\;{\rm{and}}\;\beta  > 0\;}\\
{( - 1 - g(x))/\beta ,}&{{\rm{if}}\;\;g(x) <  - (1 - \beta )\;{\rm{and}}\;\beta  > 0}
\end{array}} \right.}\\
{{\rm{where}}\;0 \le \beta  \le 1\;\;{\rm{and}}\;\;g(x) \equiv \left( {\bmod \,{\mkern 1mu} {\mkern 1mu} (x + 1,2) - 1} \right).}
\end{array}
\label{tent_map}
\end{equation}
The definition of $g(x)$ shows that it folds the input signal into an output range of $[-1,1]$. Also, the positive and negative slopes in the resulting piecewise linear map are $1/(1-\beta)$ and $-1/\beta$, respectively. For $\beta=1/2$ we get the symmetric tent map, i.e., equal positive and negative slopes. We therefore refer to $\beta$ as the asymmetry parameter. Fig.~\ref{fig:hyperchaos_map_2_plots} shows the function $f(x,\beta)$ for several values of $\beta$.

Following~\cite{Pecora1997,Carroll1998}, we define the matrix $\mathbf{A}$ as
\begin{equation}
{\bf{A}} = \left[ {\begin{array}{*{20}{c}}
a&0&b\\
0&c&1\\
1&1&0
\end{array}} \right],
\end{equation}
which is known to be hyperchaotic when $a<-1$, $b\approx 1$, and $|c|<1$. In particular, we use $a=-4/3$, $b=1$, and $c=1/3$ as default parameter values.

Lyapunov exponents (LEs) can be estimated by first finding the Jacobian $\mathbf{J}$ of the map defined in (\ref{system_eqn}). The latter is given by
\begin{equation}
\mathbf{J}=Df\cdot\mathbf{A},
\end{equation}
where $Df$ is the Jacobian of the folding function $f(\mathbf{x},\beta)$. Next, we denote $\mathbf{J_{n}}$ as the Jacobian at time step $n$. The matrix $\mathbf{Y_{n}}$ that describes evolution of the system till this time step is given by $\mathbf{Y_{n}}=\mathbf{J_{n-1}J_{n-2}}\ldots \mathbf{J_{0}}$. The multiplicative ergodic theorem~\cite{Oseledets1968} then guarantees that the following symmetric positive definite matrix (known as the Oseledec matrix) exists
\begin{equation}
\mathbf{\Lambda}=\lim_{n\rightarrow\infty}\left(\left(\mathbf{Y^{n}}\right)^{T}\cdot\mathbf{Y^{n}}\right)^{\frac{1}{2n}}.
\label{oseledec_matrix}
\end{equation}
The logarithms of the eigenvalues of $\mathbf{\Lambda}$ are the LEs. 

It is evident that $\mathbf{\Lambda}$ is the average of $\mathbf{JJ^{T}}$ along the system trajectory. It can be analytically evaluated in certain cases. Firstly, since the folding function $f(x,\beta)$ in (\ref{system_eqn}) is independently applied to each state variable, $Df$ is a diagonal matrix. Moreover, when $\beta=0$ or 1, $f(x)$ has a constant slope (1 or -1, respectively) for all $x$, so $Df$ reduces to $\pm\textbf{I}$ where $I$ is the identity matrix. Thus, in these special cases $\mathbf{J_{n}}=\pm\mathbf{A}$, i.e., is constant for all $n$. Thus, the averaging operation in (\ref{oseledec_matrix}) is not needed and each Lyapunov exponent is simply equal to $\lambda_{i} = \ln\left(\Lambda_{i}\right)/2$, where $\Lambda_{i}$ is the $i$-th eigenvalue of $\mathbf{JJ^{T}}=\mathbf{AA^{T}}$. For the default set of parameters ($a=-4/3$, $b=1$, and $c=1/3$), the resulting LE spectrum is given by (0.683, 0.302, -0.985). The presence of two positive LEs indicates hyperchaos. Moreover, in this case the sum of the LEs is zero, so the map is locally volume-preserving (VP)~\cite{Pecora1997}.

\begin{figure}[h]
\begin{center}
\includegraphics[width=0.48\textwidth]{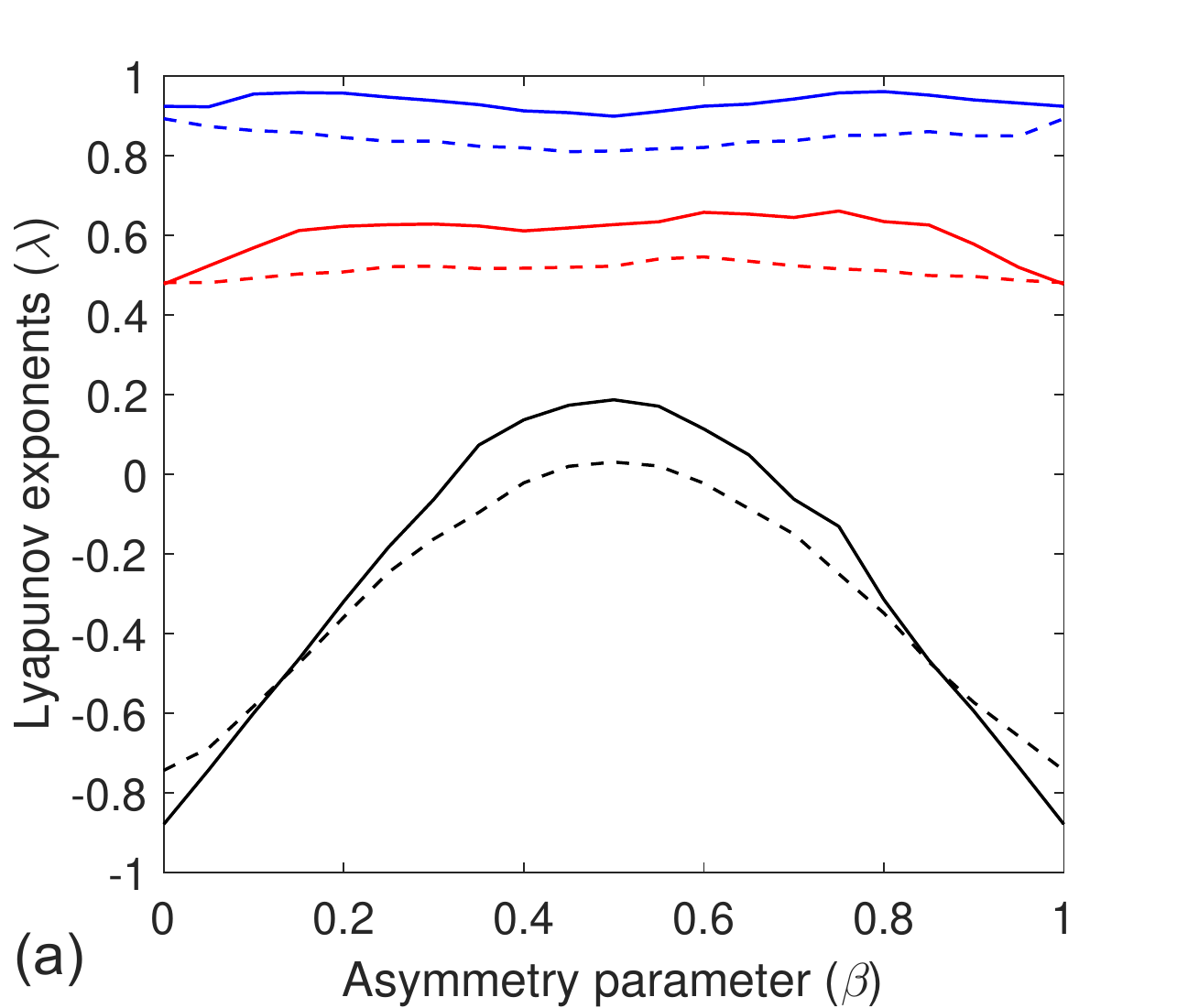}
\includegraphics[width=0.48\textwidth]{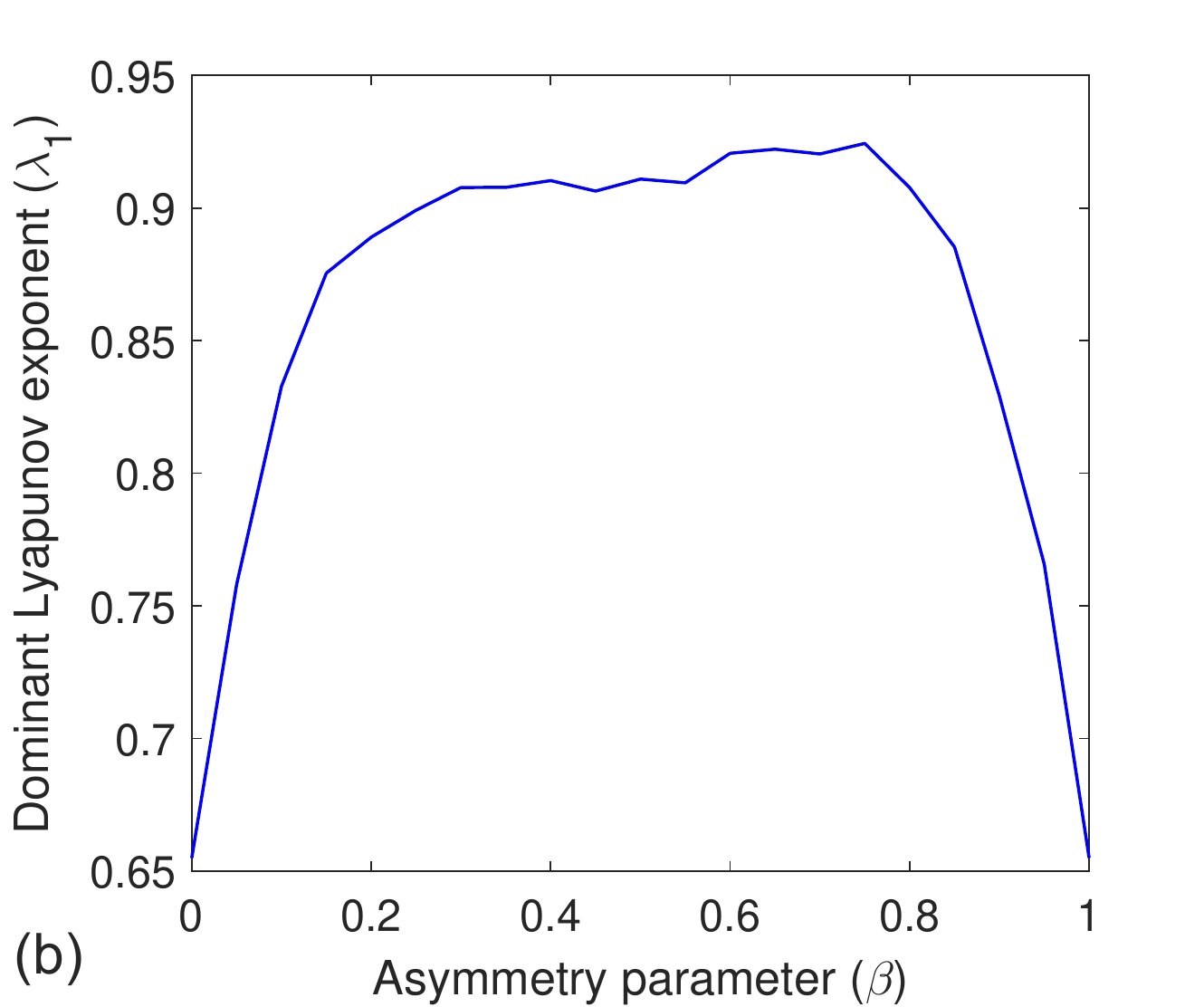}
\end{center}
\caption{(a) Simulated Lyapunov exponent spectrum of the proposed hyperchaotic system as a function of the asymmetry parameter $\beta$ using the Eckmann-Ruelle and Sano-Sawada algorithms (solid and dashed lines, respectively). (b) Simulated dominant Lyapunov exponent of the same system using the Wolf et al. algorithm.}
\label{fig:hyperchaos_map_2_lyapunov_summary}
\end{figure}

In the general case, i.e., for arbitrary values of $\beta$, the slopes of the upward and downward segments of the piecewise-linear folding function $f(x,\beta)$ are different. Specifically, $Df$ is either $1/(1-\beta)$ or $-1/\beta$, so $\mathbf{J}=Df\cdot\mathbf{A}$ can take one of two values and we have to evaluate the matrix average in (\ref{oseledec_matrix}) along the trajectory to estimate LEs. Various algorithms are available for this purpose; we used the Eckmann-Ruelle method~\cite{Eckmann1986} to numerically estimate the LE spectrum. This method uses a small neighborhood of points to estimate the local Jacobian, and then determines LEs from the eigenvalues of Jacobians estimated around the attractor. Since the folding of the attractor brings diverging orbits back together, the algorithm tends to slightly underestimate the positive exponents. For comparison, we also estimated the LE spectrum using the Sano-Sawada algorithm, which uses a similar approach~\cite{Sano1985}. The results are shown in Fig.~\ref{fig:hyperchaos_map_2_lyapunov_summary}(a) as a function of the asymmetry parameter $\beta$. For $\beta = 0$ and 1, the estimated LEs are (0.925, 0.478, -0.879) and (0.894, 0.482, -0.743) for the Eckmann-Ruelle and Sano-Sawada algorithms, respectively. These values are in good agreement with each other but somewhat larger than the theoretical ones (particularly for the first LE). Thus, the sum of the estimated LEs is positive, which corresponds to a locally volume-expanding (VE) map. Interestingly, in both cases the first two LEs are almost constant versus $\beta$, which shows that the hyperchaotic behavior is robust to the symmetry of the folding map. By contrast, the third LE is maximum for the symmetric folding map ($\beta=0.5$) and decreases strongly as $f(x,\beta)$ becomes more asymmetric.

Given the well-known numerical issues with estimating the entire LE spectrum, we also estimated the dominant LE as a function of $\beta$ using the direct algorithm proposed by Wolf et al.~\cite{Wolf1985}, which is widely used and known to be robust for a variety of dynamical systems. The results are shown in Fig.~\ref{fig:hyperchaos_map_2_lyapunov_summary}(b). They confirm the robustness of the dominant LE with respect to $\beta$ and are also in general agreement with those from the LE spectrum estimation algorithms. Moreover, the estimated value at $\beta = 0$ and 1 is 0.655, which matches the theoretical value of 0.683. Given that the various LE estimation methods are in good agreement, for the remainder of the paper we will use the Eckmann-Ruelle algorithm to estimate and compare LE spectra.

\begin{figure}[h]
\begin{center}
\includegraphics[width=0.48\textwidth]{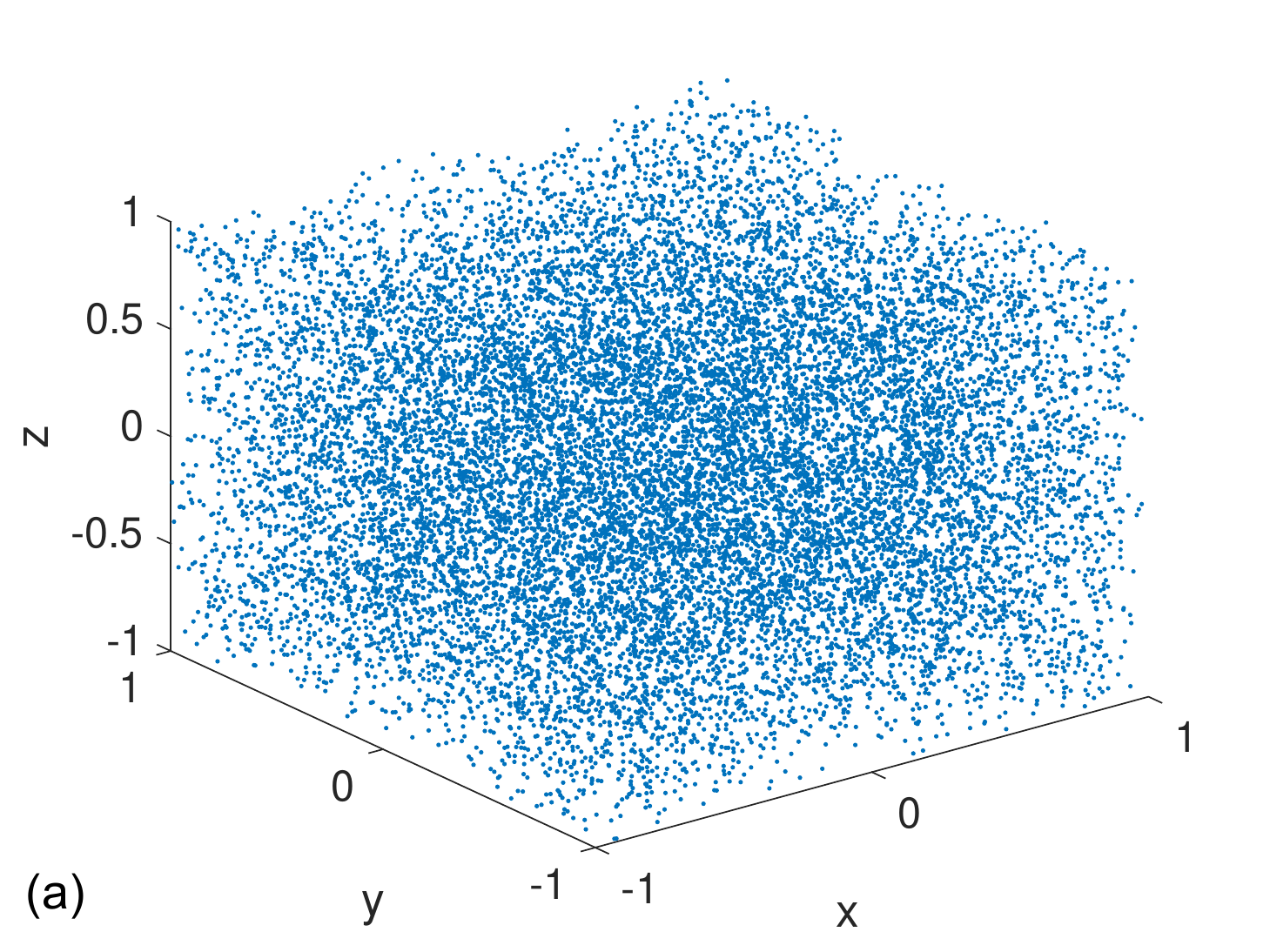}
\includegraphics[width=0.48\textwidth]{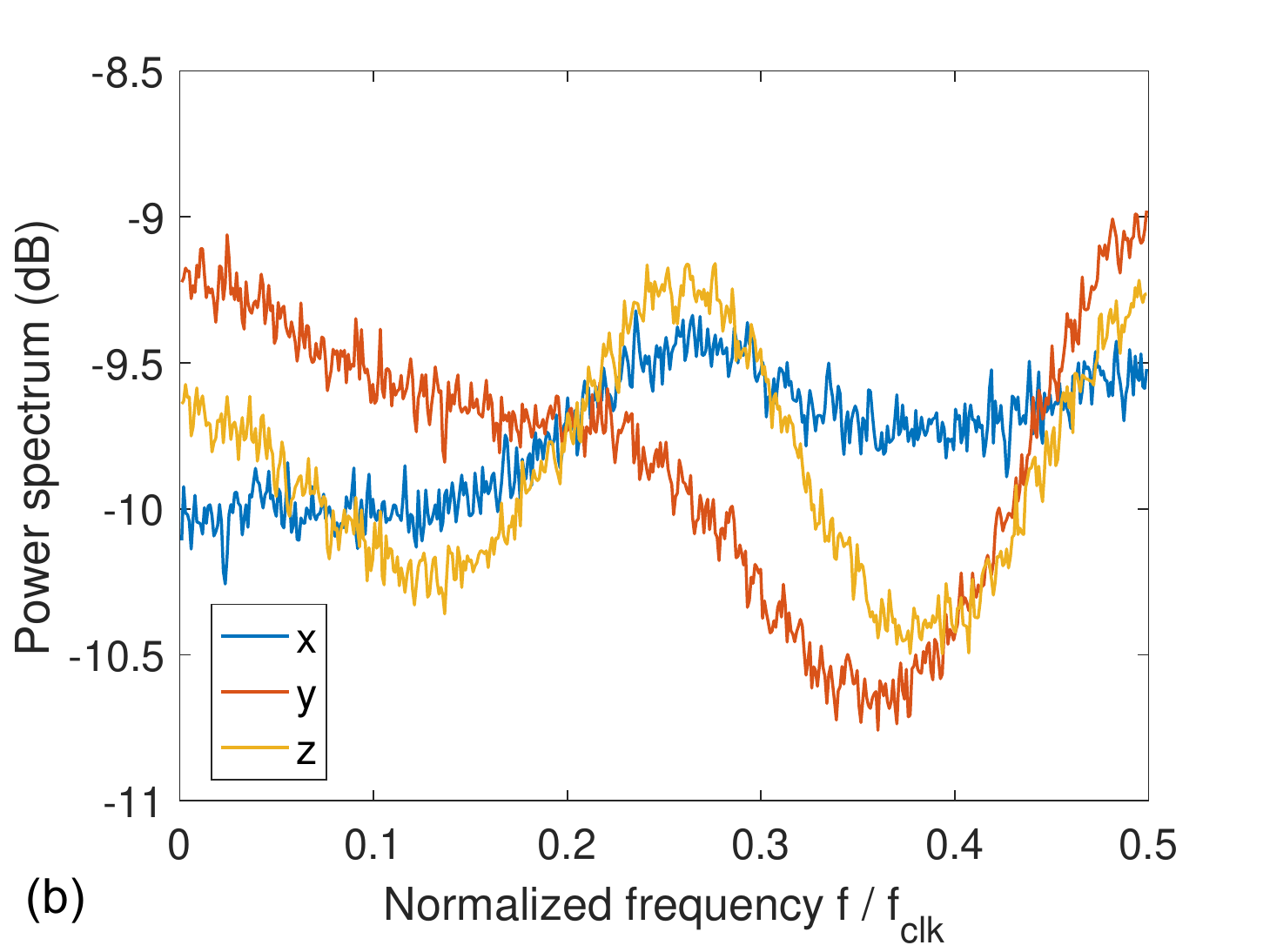}
\end{center}
\caption{Simulated properties of the hyperchaotic system for $\beta=0.5$: (a) phase space, and (b) estimated power spectra.}
\label{fig:hyperchaos_map_2_phase_psd}
\end{figure}

The simulated correlation dimension~\cite{Grassberger1983} $d_{c}$ is approximately a quadratic function of $\beta$. It increases from $2.72\pm 0.037$ (at $\beta=0$ and 1) to $2.85\pm 0.037$ (at $\beta=0.5$), where the error bars are estimated from the residuals after quadratic fitting. Since $d_{c}$ is close to 3 for any value $\beta$, we expect relatively structure-free trajectories that fill most of the available phase space. Such behavior is evident in Fig.~\ref{fig:hyperchaos_map_2_phase_psd}(a), which shows a simulated trajectory for $\beta=0.5$. The corresponding power spectra estimated using Welch's method~\cite{Welch1967} are similarly structure-free (nearly white) for all three state variables (denoted by $x$, $y$, and $z$) at frequencies up to $f_{clk}/2$ as shown in Fig.~\ref{fig:hyperchaos_map_2_phase_psd}(b). Thus, the trajectories have high entropy and short correlation times.

\section{Synchronization}
In this paper, we use the technique of synchronous substitution to synchronize two nominally-identical hyperchaotic systems denoted by transmitter (or drive) and receiver (or response), respectively. In this method, we apply an invertible transformation $T$ to the transmitter variables $\mathbf{x^{t}}$ such that $\mathbf{w}=T\left(\mathbf{x^{t}}\right)$, and then transmit only the first component of $\mathbf{w}$ (denoted by $w_{1}$), which is a scalar. At the receiver, we can recover $\mathbf{x^{t}}$ by using the inverse transformation $T^{-1}$ as long as the two systems are (nearly) synchronous such that $\mathbf{x^{r}}\approx \mathbf{x^{t}}$. To do this, we create the new variables $\mathbf{u}=T\left(\mathbf{x^{r}}\right)$ and note they should be close to $\mathbf{w}$. Thus, by using the transmitted signal $w_{1}$ for the first component and $u_{i}$ for the other components, we can create the new vector ${\bf{\tilde w}} = \{ {w_1},{u_2}, \ldots {u_n}\}  \approx {\bf{w}}$, where $n$ is the total number of state variables. The receiver now has an estimate ${\bf{\tilde x}}^{\bf{t}}=T^{-1}\left(\bf{\tilde w}\right)$ of the transmitted variables that is continuously updated by the received variable $w_{1}$. Finally, one or more components of ${\bf{\tilde x}}^{\bf{t}}$ are used to drive the response system.

The transformation $T$ must be chosen properly to ensure synchronization. In practice we use the stability of the response dynamics to chose a suitable form for $T$. Moreover, for simplicity we i) limit ourselves to linear transformations, and ii) use ${\bf{\tilde x}}^{\bf{t}}$ to create an additive feedback term in the response dynamics, as follows:
\begin{equation}
\mathbf{x^{r}_{n+1}}=f\left(\mathbf{A}\mathbf{x^{r}_{n}}+\mathbf{C}\left({\bf{\tilde x}}^{\bf{t}}-\mathbf{x^{r}}\right),\beta\right),
\end{equation}
where $\mathbf{C}$ is a coupling matrix. The coupling term can also be written as $\mathbf{C}T^{-1}\left({\bf{\tilde w}}-\mathbf{u}\right)$. In addition, we find that $\left({\bf{\tilde w}}-\mathbf{u}\right)=\left(\mathbf{K}^{T}\left(\mathbf{x^{t}}-\mathbf{x^{r}}\right),0,...0\right)$, where $\mathbf{K}$ is the first row of $T$. Considering $\mathbf{C}=\mathbf{I}$ for simplicity, the coupling term is given by $\mathbf{B\otimes K}^{T}\left(\mathbf{x^{t}}-\mathbf{x^{r}}\right)$ where $\mathbf{B}$ is the first column of $\mathbf{C}T^{-1}$ and $\otimes$ denotes the outer (or tensor) product. Thus, the coupling modifies the Jacobian of the receiver to
\begin{equation}
\mathbf{J^{\prime}}=Df\cdot\left(\mathbf{A}-\mathbf{B\otimes K}^{T}\right)=\mathbf{J}-Df\cdot\mathbf{B\otimes K}^{T}.
\end{equation}

In our case we select $B=(b,1,0)$ and $K=(\Gamma,0,1)$, such that the output variable is $w_{n}=\Gamma x_{n}+z_{n}$ where the three state variables are denoted by $x$, $y$, and $z$ for convenience. We also choose ${\tilde z}_{n}^{t}=w_{n}-\Gamma{\tilde x}_{n}^{r}$ to drive the response system, which results in the following update equations:
\begin{align}
\nonumber x_{n+1}^{r}&=f\left(ax_{n}^{r}+b{\tilde z}_{n}^{t},\beta\right),\\
y_{n+1}^{r}&=f\left(cy_{n}^{r}+{\tilde z}_{n}^{t},\beta\right),\\
\nonumber z_{n+1}^{r}&=f\left(x_{n}^{r}+y_{n}^{r},\beta\right).
\end{align}

The matrix that defines the Jacobian of this system is given by
\begin{equation}
\left( {{\bf{A}} - {\bf{B}} \otimes {{\bf{K}}^T}} \right) = \left[ {\begin{array}{*{20}{c}}
a&0&b\\
0&c&1\\
1&1&0
\end{array}} \right] - \left[ {\begin{array}{*{20}{c}}
{b\Gamma }&0&b\\
\Gamma &0&1\\
0&0&0
\end{array}} \right] = \left[ {\begin{array}{*{20}{c}}
{a - b\Gamma }&0&0\\
{ - \Gamma }&c&0\\
1&1&0
\end{array}} \right].
\end{equation}
The eigenvalues of this matrix are $(a-b\Gamma)$, $c$, and 0. Moreover, the possible values of $Df$ are $\pm 1/(1-\beta)$ and $\pm 1/\beta$, except for the special cases $\beta=0$ and 1, which result in $Df=1$ and -1, respectively. Thus, the possible eigenvalues of the receiver Jacobian are given by $\pm(a-b\Gamma,c,0)/(1-\beta)$ and $\pm(a-b\Gamma,c,0)/\beta$. The magnitudes of all eigenvalues must be $<1$ to keep the receiver system stable, i.e., maintain synchronization, which requires that
\begin{equation}
\begin{array}{l}
\beta  = 0\;{\rm{or 1}}:\quad \left| {a - b\Gamma } \right| < 1\quad {\rm{and}}\quad \left| c \right| < 1.\\
0 < \beta  \le {\textstyle{1 \over 2}}:\quad \left| {a - b\Gamma } \right| < \beta \quad {\rm{and}}\quad \left| c \right| < \beta .\\
{\textstyle{1 \over 2}} < \beta  < 1:\quad \left| {a - b\Gamma } \right| < \left( {1 - \beta } \right)\quad {\rm{and}}\quad \left| c \right| < (1 - \beta ).
\end{array}
\end{equation}
The condition on $|c|$ limits the useful values of $\beta$ to 0, 1, and the range $|c|<\beta<1-|c|$ (i.e., $1/3<\beta<2/3$ for the default case when $c=1/3$). In addition, the condition on $|a-b\Gamma|$ limits the useful range of $\Gamma$. For example, for the default case with $a=-4/3$, $b=1$, and $\beta=1/2$, it becomes $|-4/3-\Gamma|<1/2$, i.e., the useful range is $-11/6<\Gamma<-5/6$.

\begin{figure}[h]
\begin{center}
\includegraphics[width=0.45\textwidth]{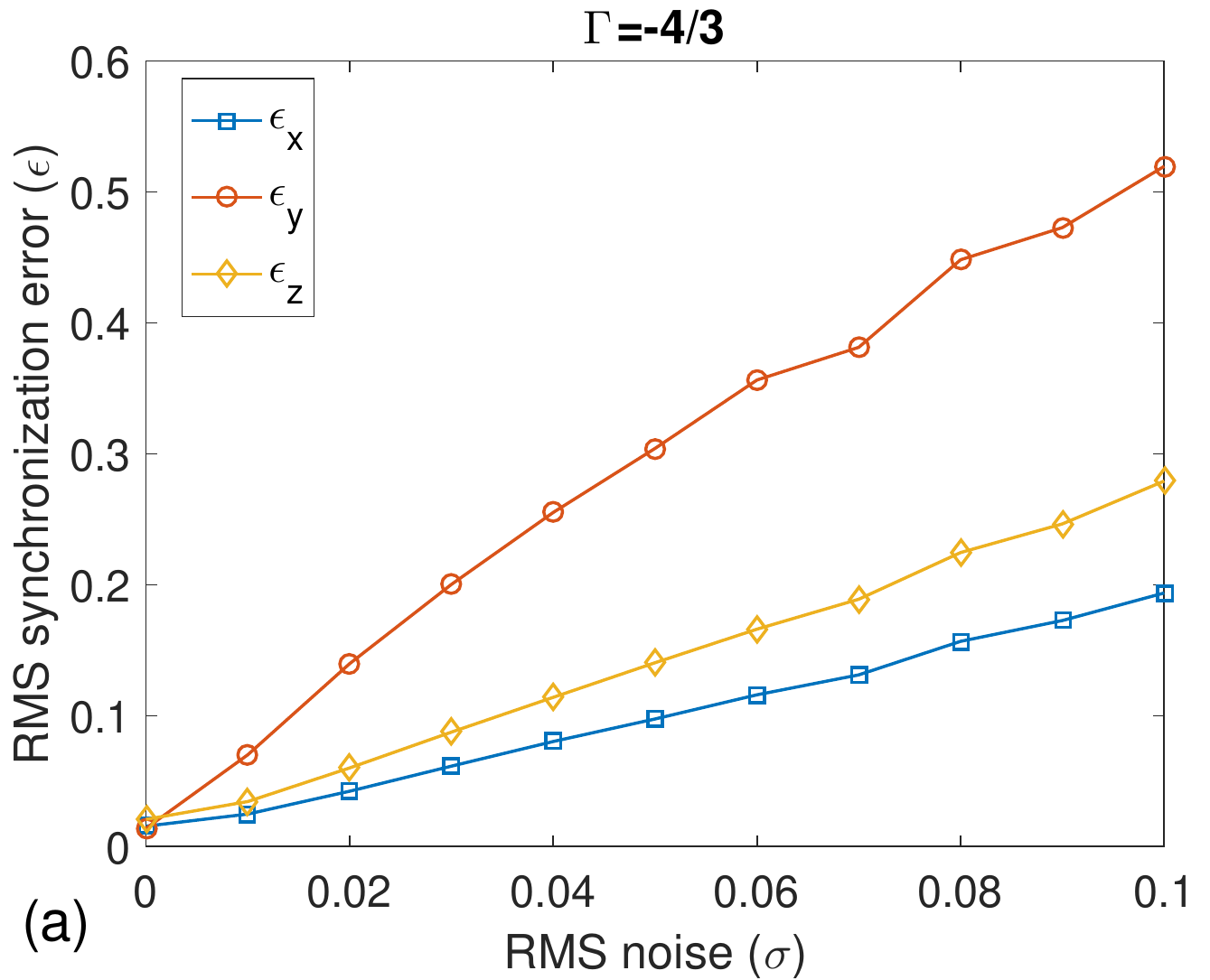}
\includegraphics[width=0.48\textwidth]{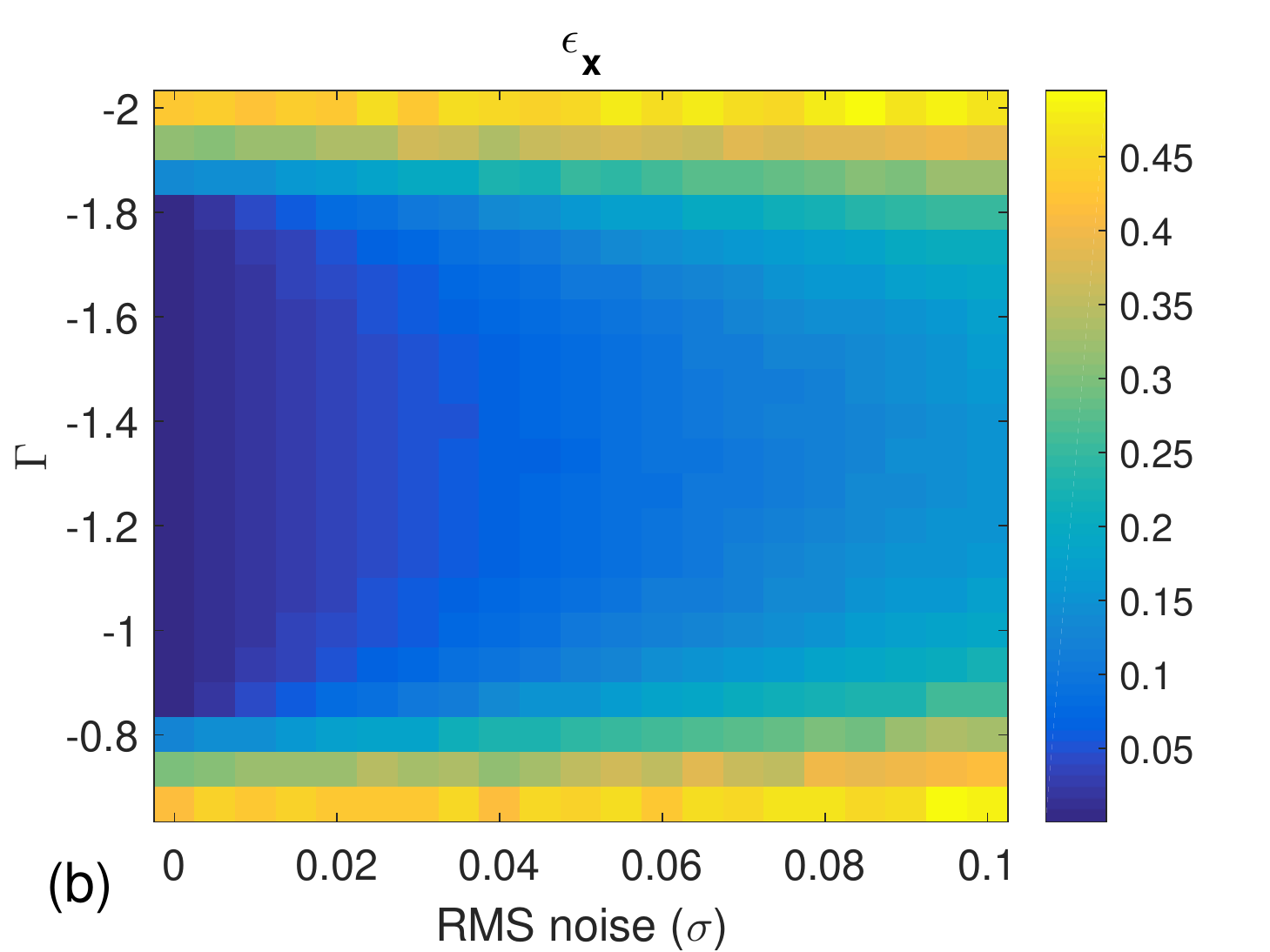}
\end{center}
\caption{Simulated synchronization performance of the hyperchaotic system for $\beta=0.5$ when white Gaussian noise of RMS value $\sigma$ is added to the drive variable $w$: (a) RMS synchronization error versus $\sigma$ for $\Gamma=a=-4/3$, and (b) RMS synchronization error for the $x$ variable versus $\sigma$ and $\Gamma$.}
\label{fig:hyperchaos_map_2_sync}
\end{figure}

To study robustness of synchronization to noise, we first implemented the drive and response systems with $\beta=1/2$ and $\Gamma=-4/3$ to ensure synchronization. We then added white Gaussian noise of standard deviation $\sigma$ to the transmitted variable $w_n$. Fig.~\ref{fig:hyperchaos_map_2_sync}(a) shows the simulated RMS synchronization error as a function of $\sigma$. Errors in all three state variables increase linearly with $\sigma$ as expected. However, there are differences in the relative sensitivity of the three variables to noise, with $x$ and $y$ having the lowest and highest sensitivity, respectively. It is interesting to note that small but non-zero synchronization errors occur even in the absence of noise (i.e., for $\sigma=0$). This is because of the non-smooth nature of the folding function $f(x,\beta)$, which can lead to sudden loss of synchronization near the fold points if one system (e.g., the drive) is folded while the other (e.g., the response) is not. This effect is strongest when the map is discontinuous ($\beta=0$ or 1) and weakest when $\beta=1/2$, so we chose the latter value for our circuit implementation.

Fig.~\ref{fig:hyperchaos_map_2_sync}(b) shows the simulated RMS synchronization error for $\beta=1/2$ as a function of $\Gamma$. The error is close to zero over the region $-11/6<\Gamma<-5/6$ (i.e., from -1.833 to -0.833) for small amounts of noise: this indicates successful synchronization as predicted by our analysis. Moreover, the error increases rapidly beyond this range, showing that synchronization fails. As in Fig.~\ref{fig:hyperchaos_map_2_sync}(a), performance degrades gradually within the synchronous region as the noise level $\sigma$ increases.

\section{Circuit Design}
\begin{figure}[h]
	\begin{center}
		\includegraphics[width=0.55\textwidth]{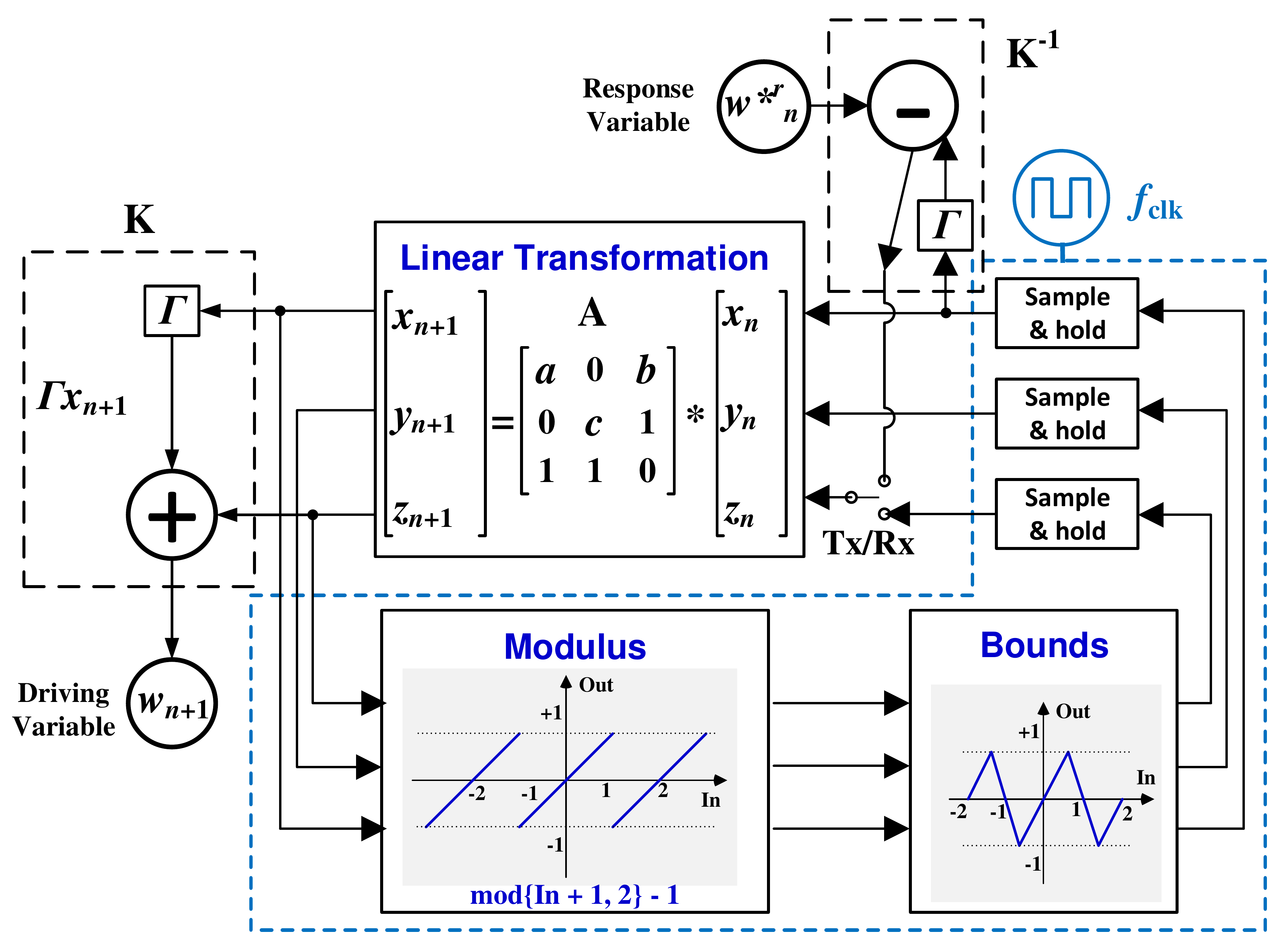}
	\end{center}
	\caption{Block diagram of the proposed discrete-time hyperchaotic system for a clock frequency of $f_{clk}$. The main blocks include the linear transformation, tent map function (modulus and bounds), and sample and hold circuits.}
	\label{fig:circuit}
\end{figure}

The proposed hyperchaotic system was implemented at the printed circuit board (PCB) level. A block diagram of the proposed hyperchaotic circuit is shown in Fig.~\ref{fig:circuit}. It consists of i) linear transformations implemented using op-amp adders and subtractors to realize $\mathbf{A}$ and $\mathbf{K}$; ii) modulus and bound operations implemented using comparators and logic gates to realize the nonlinear function $f(x,0.5)$, and iii) sample-and-hold amplifiers to realize clocked operation. A switch allows the system to operate either as a transmitter (generating the scalar output $w_{n}$) or as a receiver. All linear signal processing is performed by high-speed dual op-amps (LT1364, Linear) with a gain-bandwidth product (GBW) of 70~MHz. The default parameter values are $a=-4/3$, $b=1$, $c=1/3$, $\Gamma=-1$, and $\beta=1/2$, but these can be easily changed using resistor values on the board.

\label{sec:design}

\subsection{Nonlinear function generator}
\begin{figure}[h]
	\begin{center}
		\includegraphics[width=0.36\textwidth]{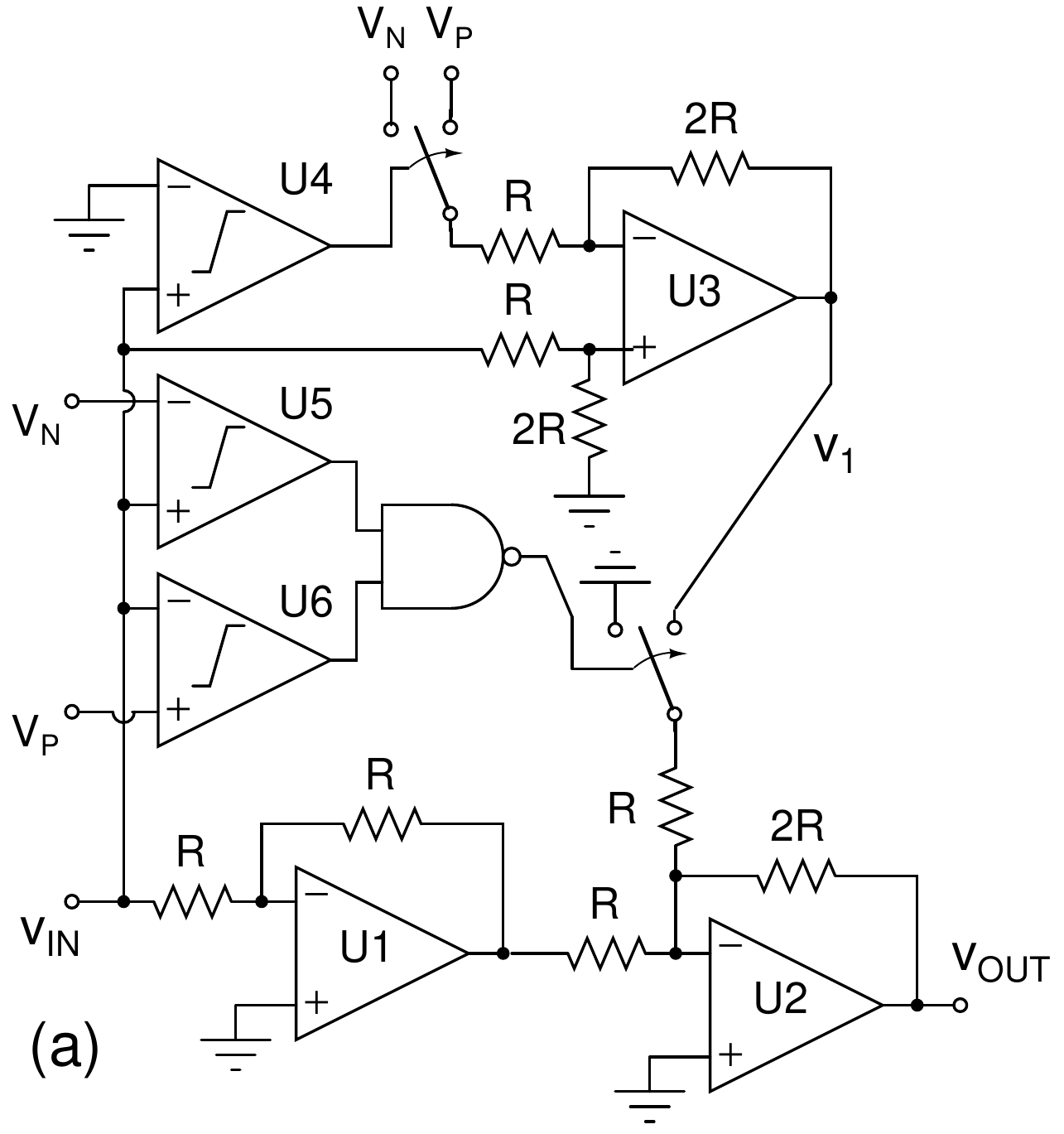}
		\includegraphics[width=0.48\textwidth]{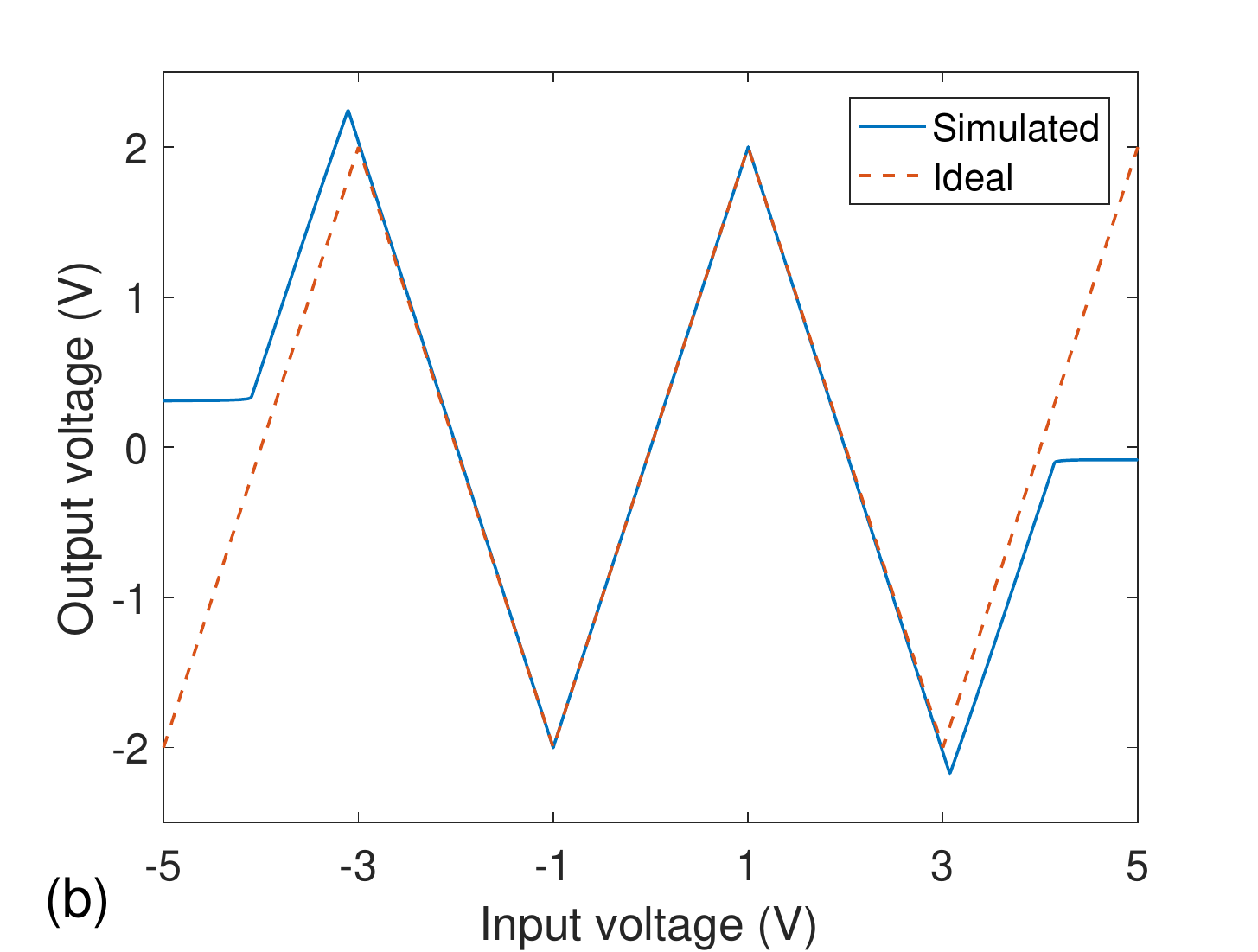}
	\end{center}
	\caption{(a) Schematic of the nonlinear function generator circuit. (b) Simulated DC transfer function of the circuit using LTspice for $V_{P}=-V_{N}=1$~V. The ideal folding function for $\beta=0.5$ is also shown for comparison.}
	\label{fig:nonlinear_gen}
\end{figure}

The nonlinear folding function generation block is shown in Fig.~\ref{fig:nonlinear_gen}(a). The voltage scale for the dynamics is set by the reference voltages $V_{P}$ and $V_{N}$, where $V_{P}=-V_{N}$ by symmetry. The op-amps U1 and U2 set $v_{IN}=v_{OUT}$ when $\left|v_{IN}\right|<V_{P}$, where $v_{IN}$ and $v_{OUT}$ are the input and output voltages, respectively. When $\left|v_{IN}\right|>V_{P}$ (i.e., for larger input voltages), the voltage $v_{1}$ (which is generated by the op-amp U3) is subtracted from $v_{OUT}$, thus implementing the folding operation. The correct region of operation is determined by the comparators U4, U5, and U6 (which compare $v_{IN}$ with $V_{N}$, ground, and $V_{P}$) and two single-pole double-throw (SPDT) analog switches.

Apart from op-amps (LT1364, Linear) and analog switches (TS12A4517, Texas Instruments), the circuit is implemented using high-speed comparators (LT1394, Linear) and two-input NOR gates (74LVC2G02, NXP) with propagation delays of $\sim 7$~ns and 1.8~ns, respectively. The unit resistance value is $R=1$~k$\Omega$, but slightly lower values (985~$\Omega$) are used to compensate for the $\sim$15~$\Omega$ on-resistance of the analog switches when necessary.

The DC transfer function of the proposed function generator was evaluated using a widely-used circuit simulator (LTspice). The results are shown in Fig.~\ref{fig:nonlinear_gen}(b) for $V_{P}=-V_{N}=1$~V. We expect to realize the transfer function $f(x,0.5)$ where $x=V/\left(V_{P}-V_{N}\right)=V/2$. The simulation closely matches this ideal function within the range $\{-3,3\}$~V, i.e., $3V_{N}$ to $3V_{P}$ or $|x|\leq 1.5$. This is sufficient for correct operation of the system, since the folding function confines each state variable within the range $\{-2,2\}$~V, i.e., $|x|\leq 1$.

\subsection{Non-overlapping clock generator}
\begin{figure}[h]
\begin{center}
\includegraphics[width=0.45\textwidth]{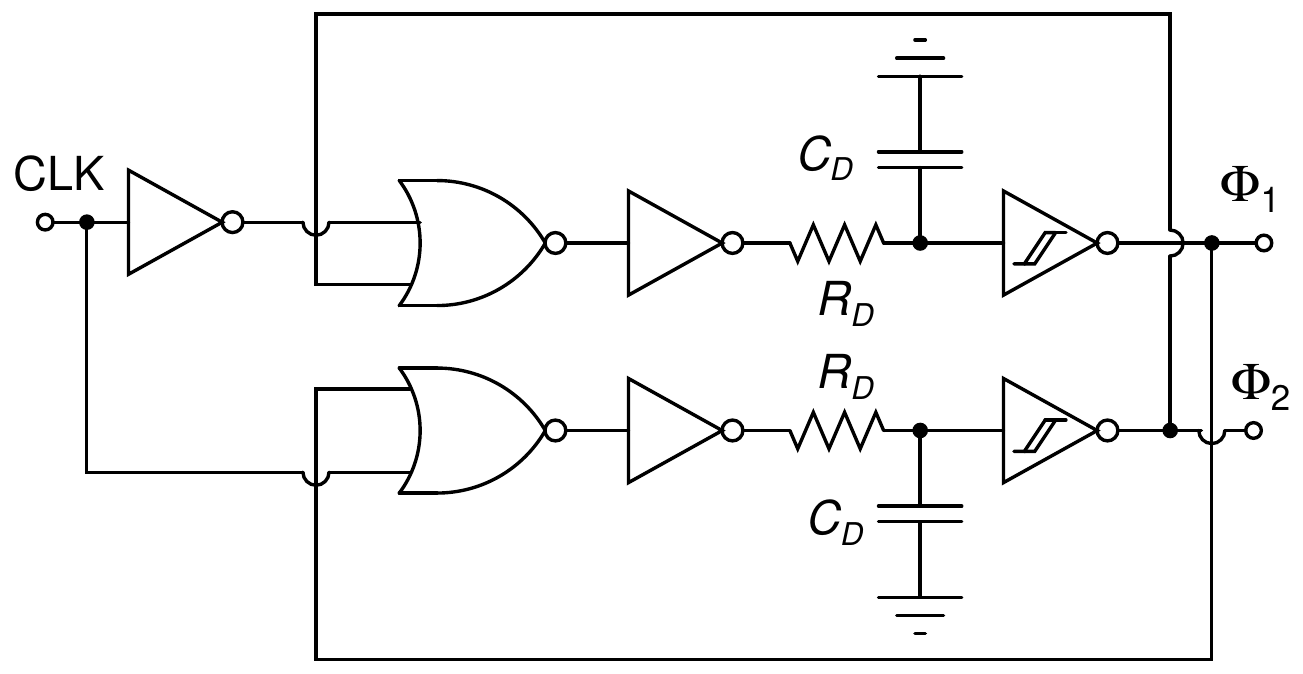}
\end{center}
\caption{Non-overlapping clock generator circuit.}
\label{fig:non_overlapping_clock_generator}
\end{figure}

The system is designed to operate using two non-overlapping clock phases, namely ``sample'' ($\phi_{1}$) and ``hold'' ($\phi_{2}$). The time delay between the two phases is set to $t_{D}$ by the non-overlapping clock generation circuit shown in Fig.~\ref{fig:non_overlapping_clock_generator}. In this circuit, $t_{D}$ is equal to the time taken by each RC circuit to charge or discharge to the threshold voltage $V_{th}$ of the Schmitt trigger. The chosen component has $V_{th}\approx 0.6V_{DD}$, so we get
\begin{equation}
V_{DD}\left(1-e^{-T_{D}/\tau_{D}}\right)=V_{th}\approx 0.6V_{DD}~\Rightarrow~T_{D}\approx\tau_{D}\ln(1/0.4)=0.92\tau_{D}.
\end{equation}

Here $\tau_{D}=R_{D}C_{D}$ is the time constant of the RC circuit. Thus, the duration of each phase (i.e., the sample and hold times) is $T=1/\left(2f_{clk}\right)-T_{D}$ where $f_{clk}$ is the clock frequency. In our case, $R_{D}=1$~k$\Omega$ and $C_{D}=100$~pF, resulting in $T_{D}\approx 92$~ns.

\subsection{Sample-and-hold}
\begin{figure}[h]
\begin{center}
\includegraphics[width=0.35\textwidth]{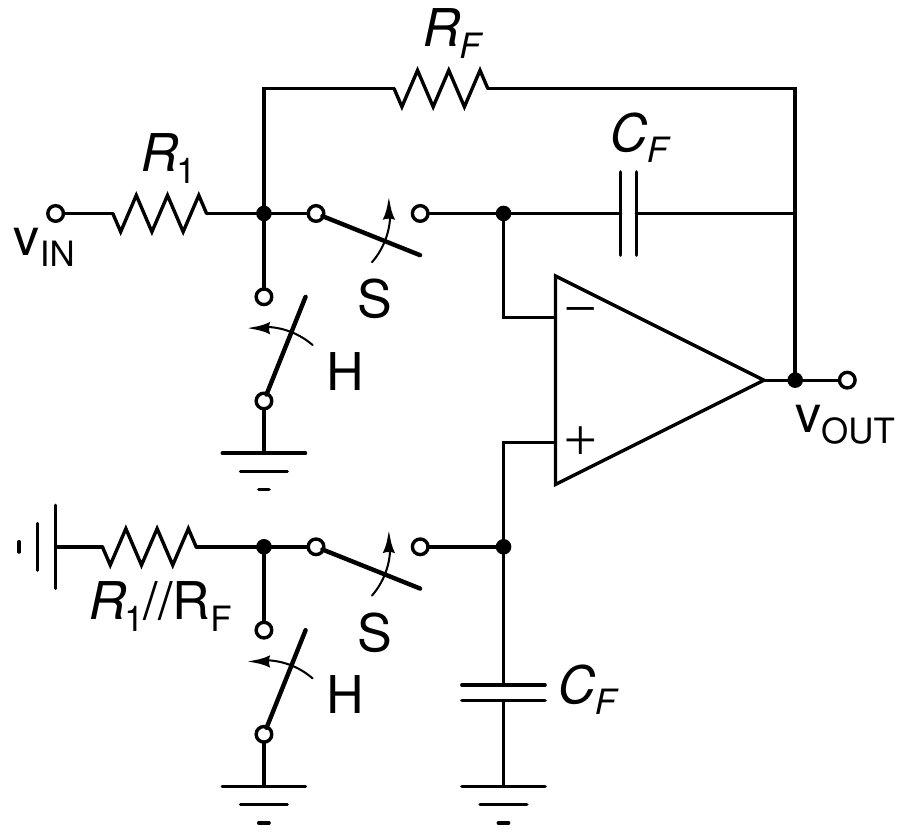}
\end{center}
\caption{Closed-loop sample-and-hold amplifier (SHA) circuit using differential switching.`S' and `H' denote the sample and hold phases of the clock, respectively.}
\label{fig:sha}
\end{figure}

Clocked analog delay elements are used for iterating the state variables. Each delay is implemented using two sample-and-hold amplifiers (SHAs) that are connected in series and clocked on opposite phases. Closed-loop SHAs were used to ensure high DC accuracy. In addition, differential switching is used to minimize the effects of charge injection from the switches, as shown in Fig.~\ref{fig:sha}.

The transfer function (TF) of each SHA in ``hold'' mode is given by
\begin{equation}
H(s)=\frac{A}{s\tau+1},~A\equiv -\left(\frac{R_{F}}{R_{1}}\right),~\tau\equiv R_{F}C_{F}.
\end{equation}

This is the TF of a first-order low-pass filter (LPF) with a -3~dB cutoff frequency of $1/\tau$. In our case $R_{F}=560$~$\Omega$ and $C_{F}=220$~pF, resulting in $\tau=123.2$~ns.

\subsection{Effects of incomplete settling}
Since each SHA behaves as a first-order LPF, its time-domain step response is an exponential function with a time constant of $\tau$, resulting in the following step response:
\begin{equation}
x(t) = x(0) + \left[ {x(\infty ) - x(0)} \right]\left( {1 - {e^{ - t/\tau }}} \right),
\end{equation}
where $x(0)$ and $x(\infty)$ are the initial and final output values, respectively. Thus, each SHA in the delay element attenuates the input step $\left(x(\infty)-x(0)\right)$ by a factor of $\left(1-e^{-T/\tau}\right)$, where $T$ is the duration of each phase ($\phi_{1}$ or $\phi_{2}$). The delay element contains two cascaded SHAs, resulting in an overall attenuation of $\left(1-e^{-T/\tau}\right)^{2}$. As a result, the system dynamics can be written as 
\begin{align}
\nonumber {{\bf{x}}_{n + 1}} &= {{\bf{x}}_n} + \left[ {f\left( {{\bf{A}}{{\bf{x}}_n},\beta } \right) - {{\bf{x}}_n}} \right]\left( {1 - {e^{ - T/\tau }}} \right)^{2}\\
 &= f\left( {{\bf{A}}{{\bf{x}}_n},\beta } \right)\left( {1 - {e^{ - T/\tau }}} \right)^{2} + {{\bf{x}}_n}{\left(2-e^{-T/\tau}\right)e^{ - T/\tau }}.
\label{hyperchaos_map_settle}
\end{align}

The second term corresponds to an exponential decay $e^{-nT/\tau}$, which has a stable fixed point at the origin.

\section{Simulations}
\label{sec:sim}
\begin{figure}[h]
\begin{center}
\includegraphics[width=0.48\textwidth]{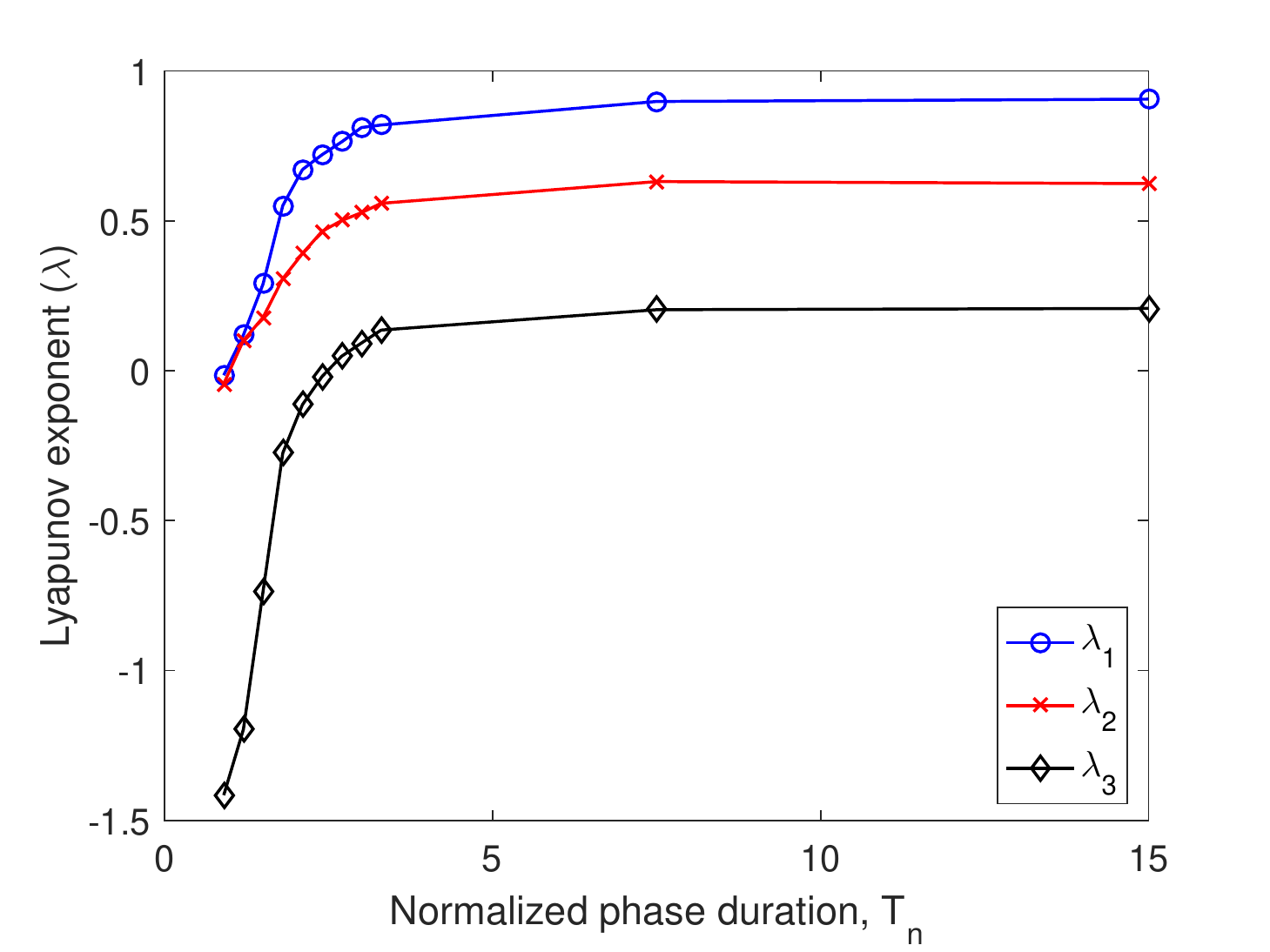}
\end{center}
\caption{Simulated LE spectrum of the non-ideal system dynamics as a function of the normalized hold time $T_{n}=T/\tau$ for $\beta=0.5$.}
\label{fig:hyperchaos_map_2_lyapunov}
\end{figure}

Fig.~\ref{fig:hyperchaos_map_2_lyapunov} shows the simulated LE spectrum of the non-ideal system dynamics defined by (\ref{hyperchaos_map_settle}) versus the normalized hold time $T_{n}=T/\tau$. The LEs monotonically decrease as $T_{n}$ decreases, i.e., settling becomes more and more incomplete. This result shows that incomplete settling has a strong damping effect on the system dynamics, as expected. Eventually the hyperchaotic behavior is lost: all three LEs become negative for $T_{n}<1$.

The proposed circuit implementation of the hyperchaotic system was simulated in LTspice using manufacturer-provided device models for various values of $f_{clk}$, i.e., $T_{n}$. The results were in good agreement with the theoretical system dynamics, but are not discussed further to save space.

\section{Experimental Results}
\label{sec:expt}
\begin{figure}[h]
\begin{center}
\includegraphics[width=0.48\textwidth]{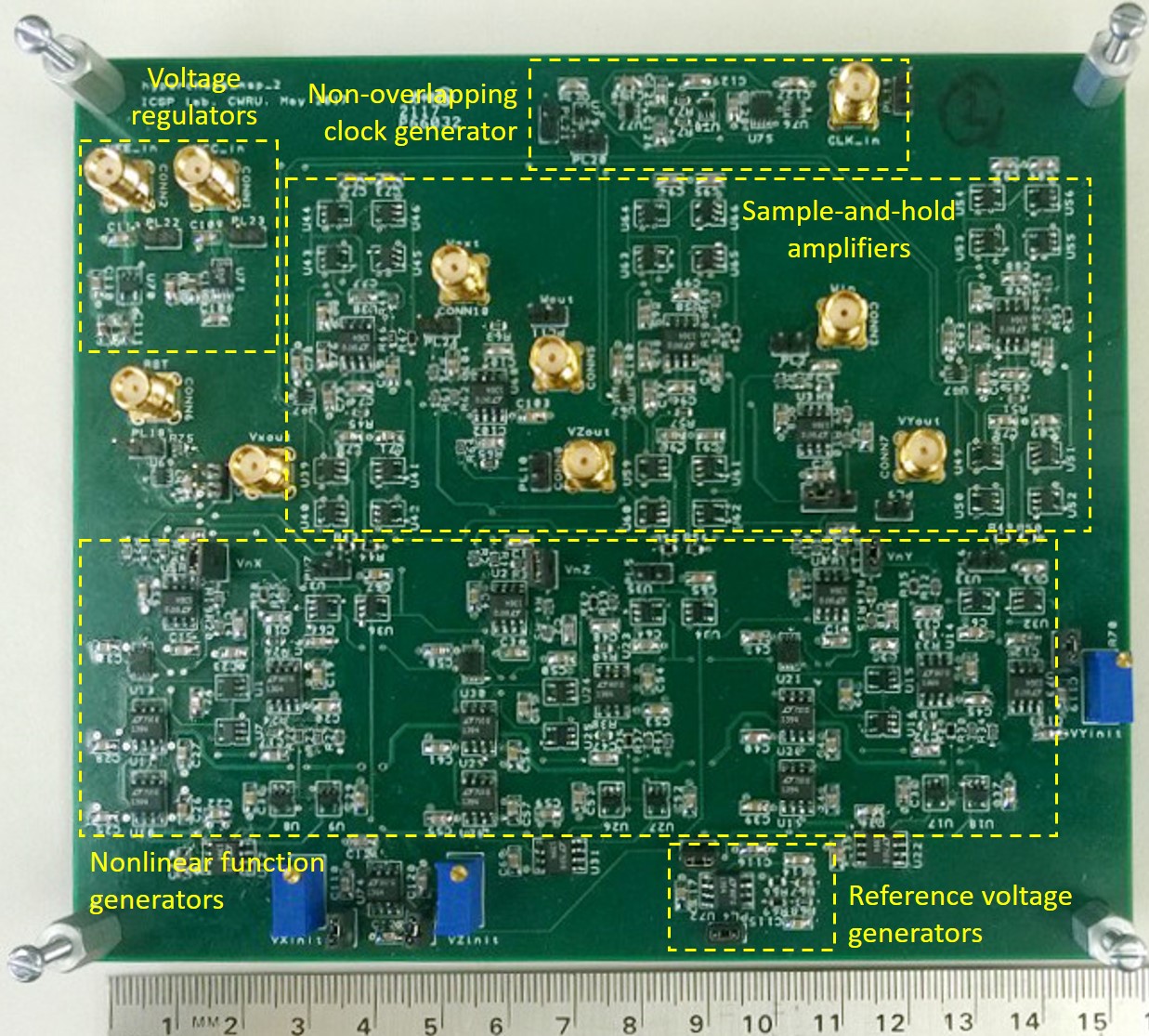}
\end{center}
\caption{Photograph of a single hyperchaotic system implemented on a PCB. Important blocks are labeled.}
\label{fig:board_photo}
\end{figure}

Fig.~\ref{fig:board_photo} shows a photograph of a single hyperchaotic system implemented on a 4-layer printed circuit board (PCB) that measures 15.3~$\times$~12.5~cm in size. Two of these PCBs were populated using nominally identical parameters in order to study synchronization and data communication applications.

Table~\ref{tab:param_values} summarizes the measured electrical parameters of the two hyperchaotic systems that were implemented. Here $V_{CC}$ and $V_{EE}$ are the on-board positive and negative power supply voltages, and $I_{CC}$ and $I_{EE}$ are the average currents drawn from them. The values of $V_{CC}$ and $V_{EE}$ differ slightly from the theoretical values of $\pm 5$~V because they are set by individual LDOs. This results in $\sim$1\% errors in the values of $V_{P}$ and $V_{N}$ (which are derived from $V_{CC}$ and $V_{EE}$ using voltage dividers) and thus the shape of the folding map.
\begin{table}[tbh]
\centering
\tbl{Measured electrical parameters of two hyperchaotic systems.}
{\begin{tabular}{c|c|c}
\hline
\textbf{Parameter} & \textbf{Board 1} & \textbf{Board 2}\\
\hline
$V_{CC}$ (V) & 5.004 & 4.971 \\
$I_{CC}$ (mA) & 244 & 240 \\
$V_{EE}$ (V) & -4.975 & -4.966 \\
$I_{EE}$ (mA) & -185 & -182 \\
$V_{P}$ (V) & 0.994 & 0.988\\
$V_{N}$ (V) & -0.988 & -0.986 \\
\hline
\end{tabular}}
\label{tab:param_values}
\end{table}

\subsection{Basic Operation}
\begin{figure}[h]
\begin{center}
\includegraphics[width=0.48\textwidth]{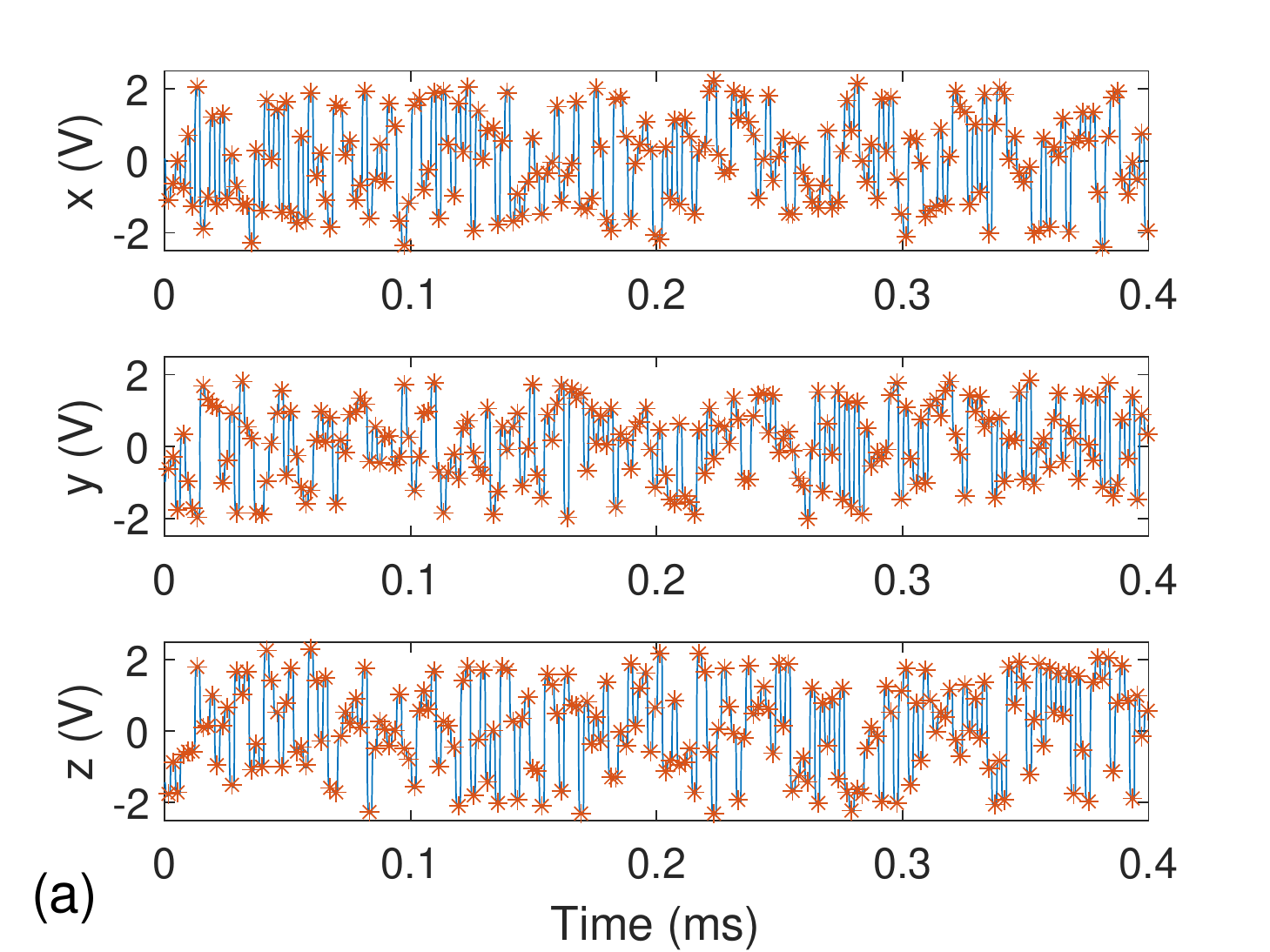}
\includegraphics[width=0.48\textwidth]{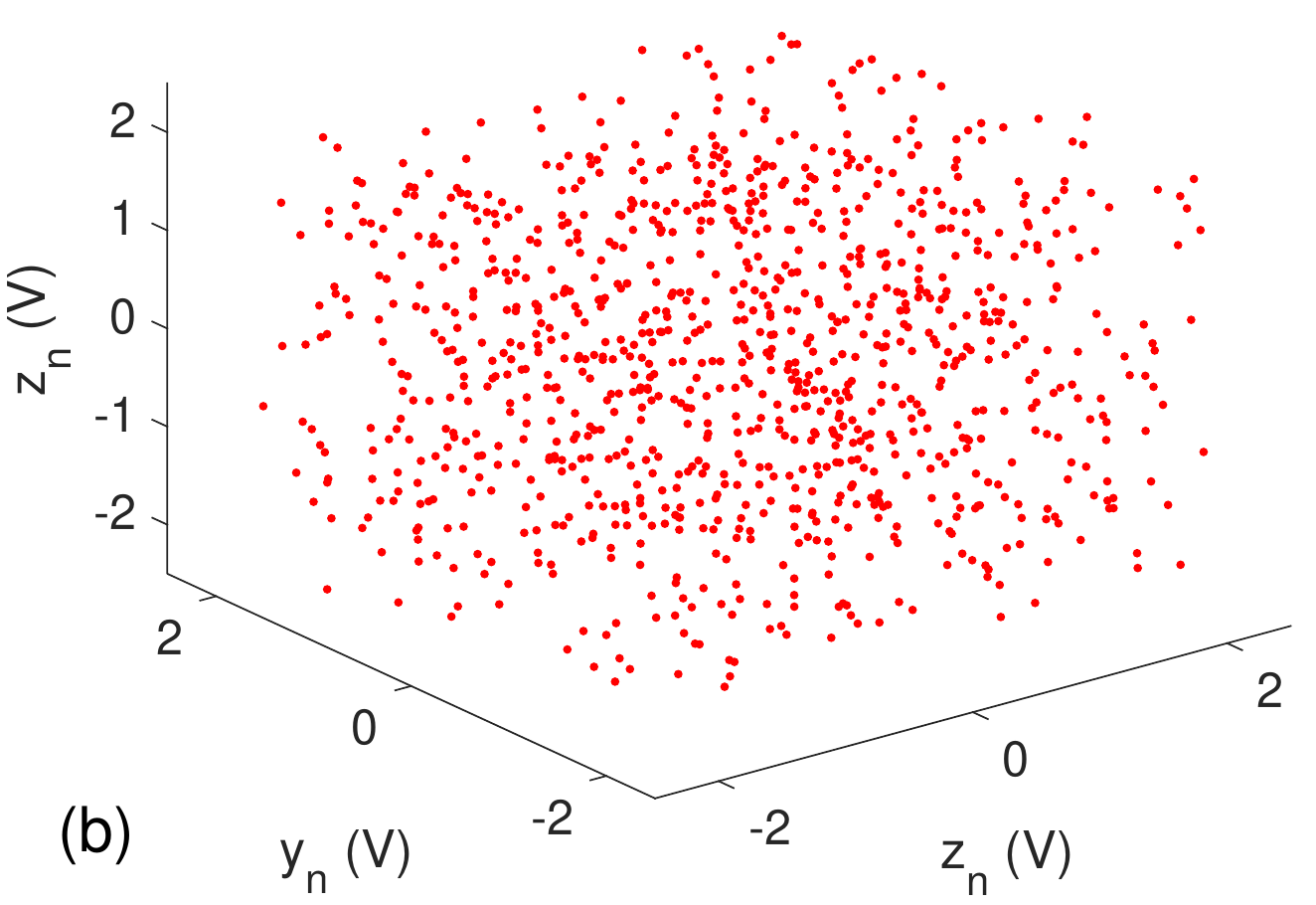}
\end{center}
\caption{Typical outputs measured at a clock frequency of $f_{clk}=0.5$~MHz: (a) time-domain waveforms, with estimated map voltages $x_{n}$, $y_{n}$, and $z_{n}$ shown as asterisks; (b) phase space of the experimentally-observed map (1000 points).}
\label{fig:typical_outputs}
\end{figure}

Both boards were tested and found to function as expected. Fig.~\ref{fig:typical_outputs} shows typical outputs from one board measured at a clock frequency of $f_{clk}=0.5$~MHz, for which $T/\tau\approx 7.37$ and incomplete settling effects are negligible. Fig.~\ref{fig:typical_outputs}(a) shows time-domain signals captured with an oscilloscope. The average voltage  within the second half of each clock period was estimated for all three variables during post-processing and stored as the discrete-time map voltages $x_{n}$, $y_{n}$, and $z_{n}$. The phase space defined by these variables is shown in Fig.~\ref{fig:typical_outputs}(b). It is in good agreement with the simulated phase space shown in Fig.~\ref{fig:hyperchaos_map_2_phase_psd}(a).

\begin{figure}[h]
\begin{center}
\includegraphics[width=0.48\textwidth]{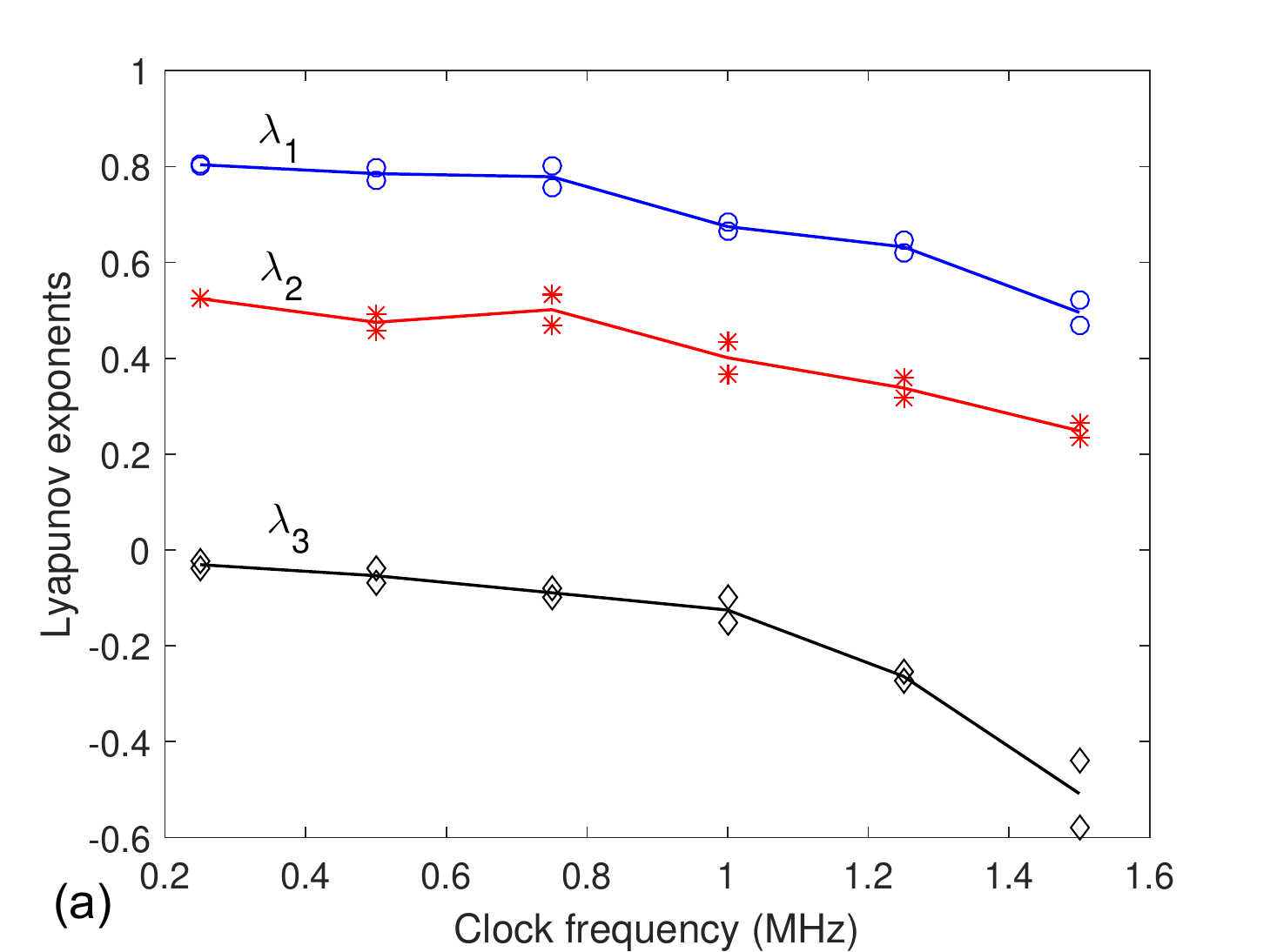}
\includegraphics[width=0.48\textwidth]{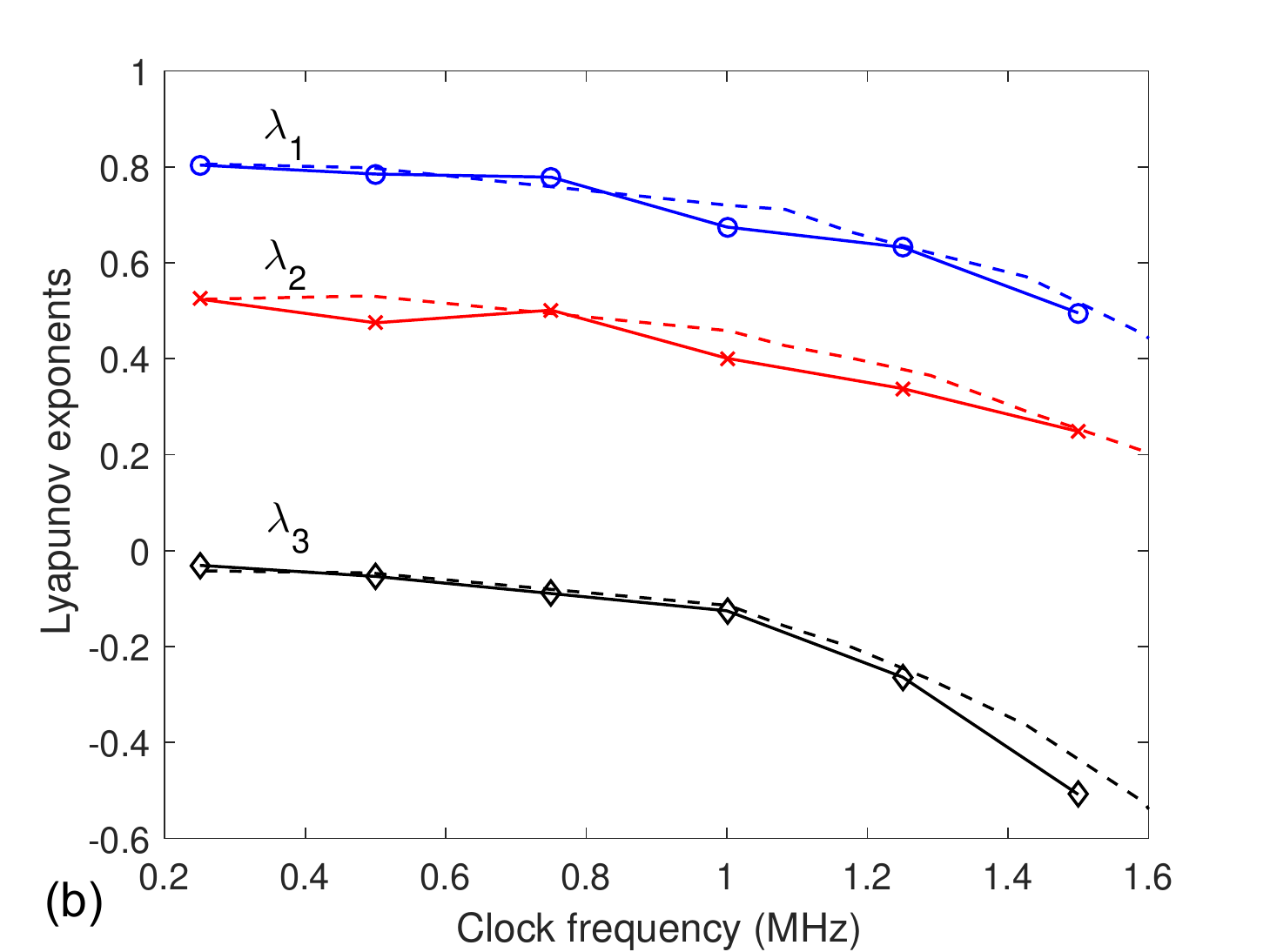}
\end{center}
\caption{(a) Measured Lyapunov spectra from two boards as a function of $f_{clk}$. Solid lines denote the average of two boards. (b) Comparison of Lyapunov spectra: measured (solid lines) and simulated (dashed lines). The measured values are the averages of two boards. Offsets of 0.1, 0.1, and 0.25, respectively have been subtracted from the simulated exponents to highlight the effect of clock frequency (i.e., phase duration) on the dynamics.}
\label{fig:lyapunov_fclk}
\end{figure}

Fig.~\ref{fig:lyapunov_fclk}(a) summarizes the Lyapunov spectra estimated from the experimental outputs of the two boards as a function of the clock frequency $f_{clk}$. The dynamics of the two maps are well-matched and clearly hyperchaotic, with two positive LEs as expected. The effects of incomplete settling are visible as a gradual decrease of all three LEs as $f_{clk}$ increases. Fig.~\ref{fig:lyapunov_fclk}(b) compares these experimental results with theoretical spectra obtained by numerically simulating the system dynamics in Eqn.~\ref{hyperchaos_map_settle}, which include incomplete settling effects. The two sets of LEs are in excellent agreement with each other once small fixed offsets are subtracted from the theoretical LE values. Thus, we conclude that Eqn.~\ref{hyperchaos_map_settle} accurately models system dynamics in the presence of incomplete settling.

\subsection{Synchronization}
\begin{figure}[h]
\begin{center}
\includegraphics[width=0.96\textwidth]{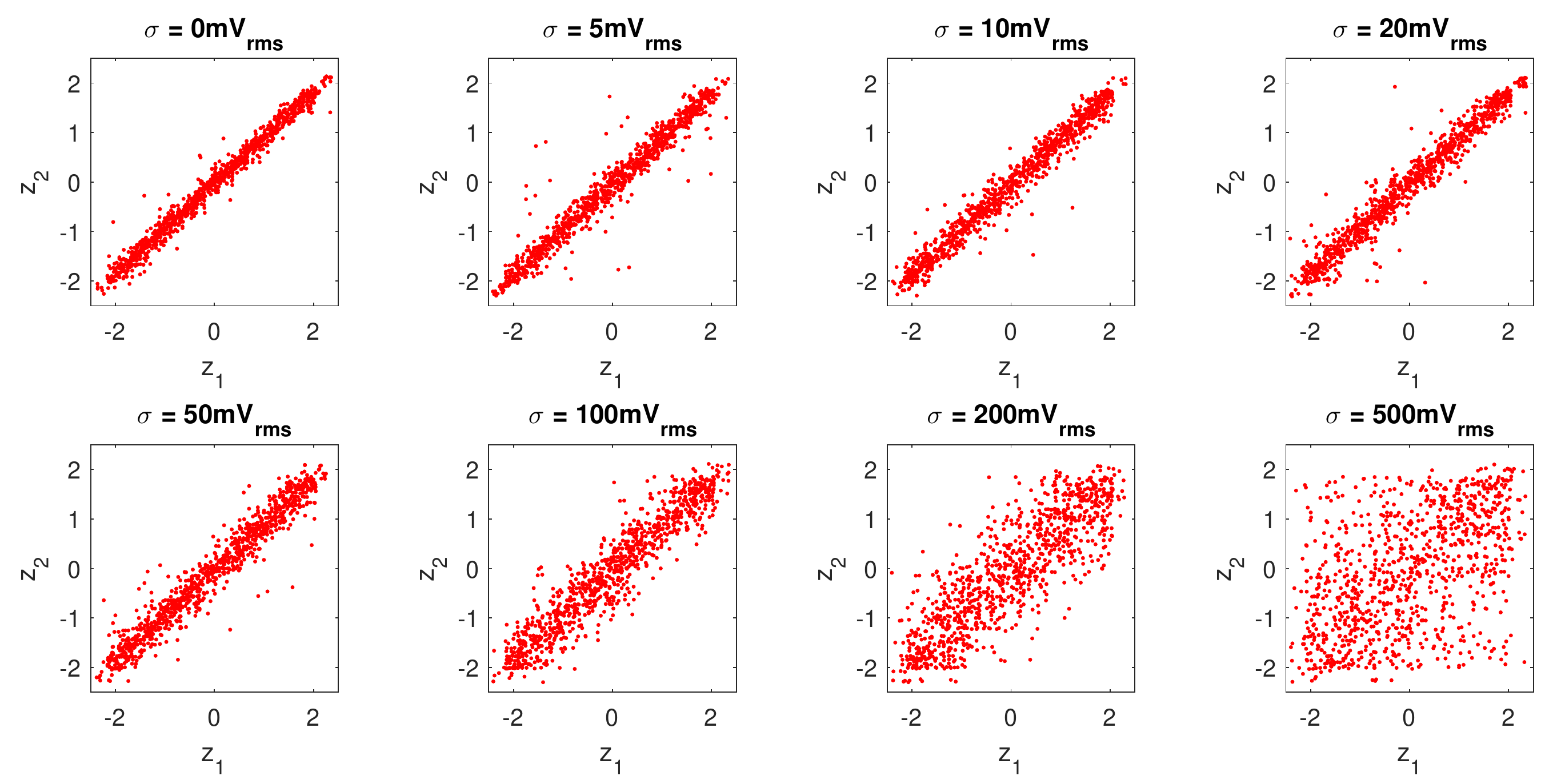}
\end{center}
\caption{Synchronization in the presence of additive white Gaussian noise (AWGN) for $f_{clk}=0.5$~MHz.}
\label{fig:sync_noise_1}
\end{figure}

\begin{figure}[h]
\begin{center}
\includegraphics[width=0.43\textwidth]{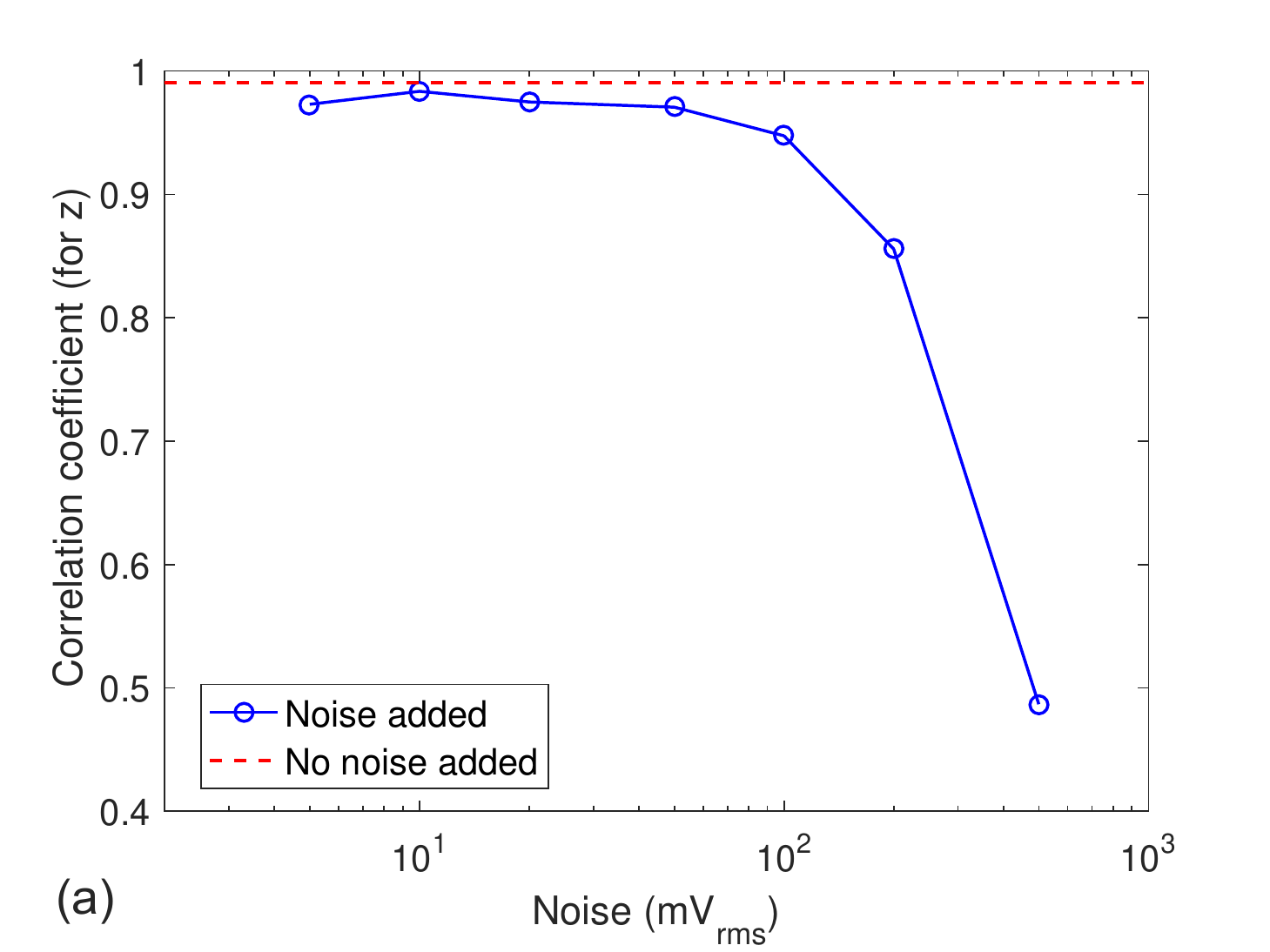}
\includegraphics[width=0.48\textwidth]{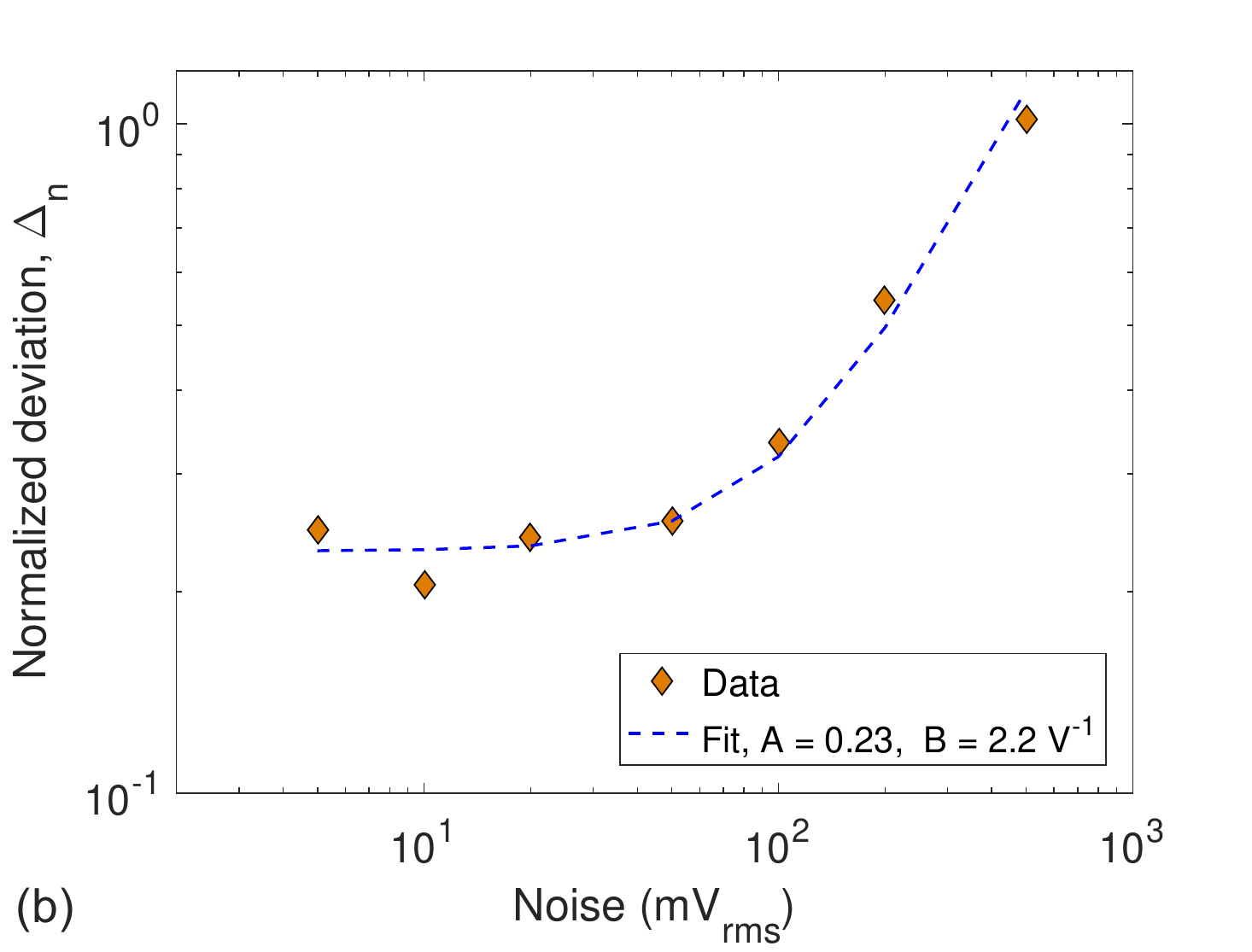}
\end{center}
\caption{Estimated (a) correlation coefficient, and (b) normalized deviation between the synchronized systems as a function of noise level for $f_{clk}=0.5$~MHz.}
\label{fig:sync_noise_2}
\end{figure}

In order to characterize synchronization performance versus noise, one board was configured as a transmitter and the other as a receiver. The two used synchronized clocks, which in practical applications can be generated by a separate clock recovery circuit such as a phase-locked loop (not discussed here). We then added white Gaussian noise with various standard deviations $\sigma$ (from a signal generator) to the output $w^{t}$ of the drive board. Fig.~\ref{fig:sync_noise_1} summarizes the experimental synchronization performance of one of the state variables ($z$) for several values of $\sigma$. As expected, synchronization gradually degrades as the noise level increases. These results are summarized in Fig.~\ref{fig:sync_noise_2}(a), which shows the measured correlation coefficient between drive and response variables ($z_{1}$ and $z_{2}$, respectively) versus $\sigma$. The former remains nearly constant for $\sigma<50$~mV before decreasing sharply, showing that the synchronization process is robust to significant amounts of additive noise.

Fig.~\ref{fig:sync_noise_2}(b) summarizes the average normalized deviation $\Delta_n$ measured between the synchronized systems as a function of noise level $\sigma$. Here $\Delta_n$ is defined as
\begin{equation}
{\Delta _n} = \frac{{{{\left[ {\left\langle {{{\left| {{z_1} - {z_2}} \right|}^2}} \right\rangle } \right]}^{1/2}}}}{{{\sigma _{{z_1}}}{\sigma _{{z_2}}}}},
\end{equation}
where $\left\langle~\right\rangle$ denotes time average, and $\sigma_{z_{1}}$ and $\sigma_{z_{2}}$ are the standard deviations of $z_{1}$ and $z_{2}$,  respectively. The results are well-fit by an expression of the form
\begin{equation}
\Delta_n = \left[A^2+(\sigma B)^2\right]^{1/2},
\end{equation}
where $A$ and $B$ are constants that are independent of noise level. This result is similar to that found earlier for synchronization of chaotic systems in the presence of additive noise and drift~\cite{Brown1994}. Here $A$ is due to the non-smooth folding function and modeling errors (e.g., deviations between the parameters of the two systems due to component tolerances), while $B$ is a function of both coupling strength and autocorrelation statistics of the noise.

\subsection{Communication via chaotic masking}
We present a private digital data transmission framework via two copies of the proposed hyperchaotic system. The systems are inherently hyperchaotic but can be synchronized by a scalar transmitted signal (which can contain a significant amount of noise, see Fig.~\ref{fig:sync_noise_2}(a)). Thus, we can mix our digital information over the transmitted dynamical variable and drive the receiver with the transformed mixed variable while maintaining synchronization; essentially, we expect the data to behave like additive noise and not disrupt synchronization as long as its amplitude is not too large and circuit parameters are sufficiently matched.

\begin{figure}[h]
\begin{center}
\includegraphics[width=0.7\textwidth]{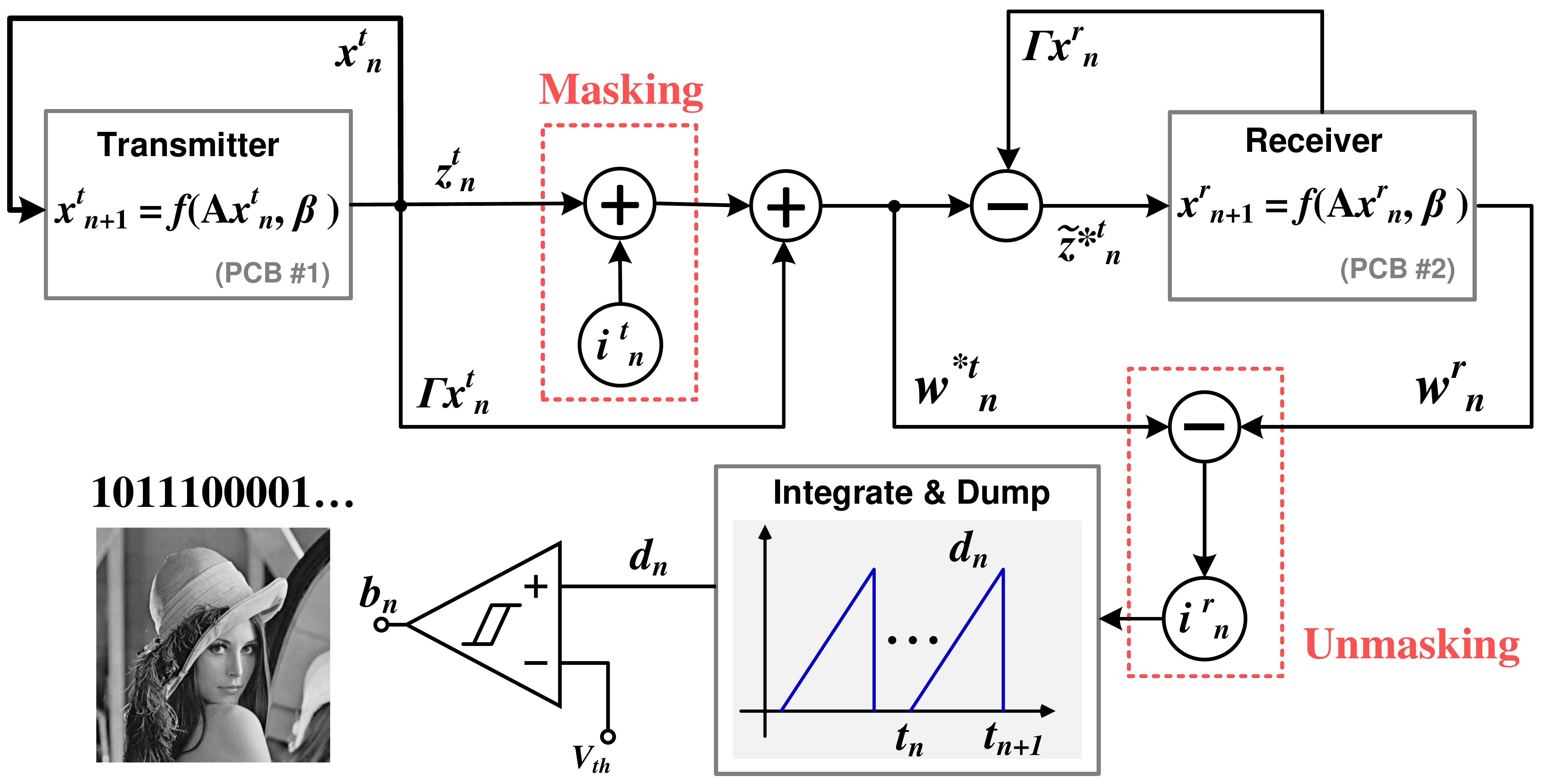}
\end{center}
\caption{Flowchart for digital communication via chaotic masking and unmasking.}
\label{fig:data_scheme}
\end{figure}

Fig.~\ref{fig:data_scheme} schematically illustrates the process of digital communication via chaotic masking and unmasking. Specifically, we represent the state functions of the transmitter and receiver chaotic system, respectively, as follows:
\begin{eqnarray}
\mathbf{x}^{t}_{n+1} = f(\mathbf{A}\mathbf{x}^{t}_{n},\beta)\\
\mathbf{x}^{t}_{n+1} = f(\mathbf{A}\mathbf{x}^{r}_{n},\beta)
\label{tx_rx}
\end{eqnarray}
where we denote `$t$' as the transmitter and `$r$' as the receiver. We choose the third variable $z^{t}_{n}$ to mix with digital information $i^{t}_{n}$ and then form the new output variable $w^{*t}_{n}$, as shown in Fig.~\ref{fig:data_scheme}. The linearly transformed driving variable $w^{*t}_{n}$ is thus defined as follows:
\begin{equation}
w^{*t}_{n}=z^{*t}_{n}+\Gamma x^{t}_{n}
\label{drive_variable}
\end{equation}
where $z^{*t}_{n}$ is the third variable after mixing, and $\Gamma$ is set to $a+1/3=-1$ to ensure that the receiver $x^{r}-y^{r}$ subsystem is stable and has low synchronization error, as described earlier. Using the synchronization substitution technique~\cite{Carroll1998}, only $\tilde{z}^{*t}_{n}=w^{*t}_{n}-\Gamma x^{r}_{n}$ is implemented as the driving variable (in place of $z^{r}_{n}$) for the receiver. When the systems are in synchrony, we have $z^{*t}_{n+1}=z^{r}_{n+1}$, the value of the third component before information mixing. Therefore, the signal can be retrieved as $i^{r}_{n}=i^{t}_{n}$ at the next time step when in synchronization. In our case, we simply mix the digital information and chaotic driving variables using addition. This simplifies chaotic unmasking to a subtraction operation: $i^{r}_{n} = z^{r}_{n+1}-z^{*t}_{n+1}$, which is equivalently $i^{r}_{n} = w^{r}_{n} - w^{*t}_{n}$.

\begin{figure}[h]
	\begin{center}
     \includegraphics[width=0.46\textwidth]{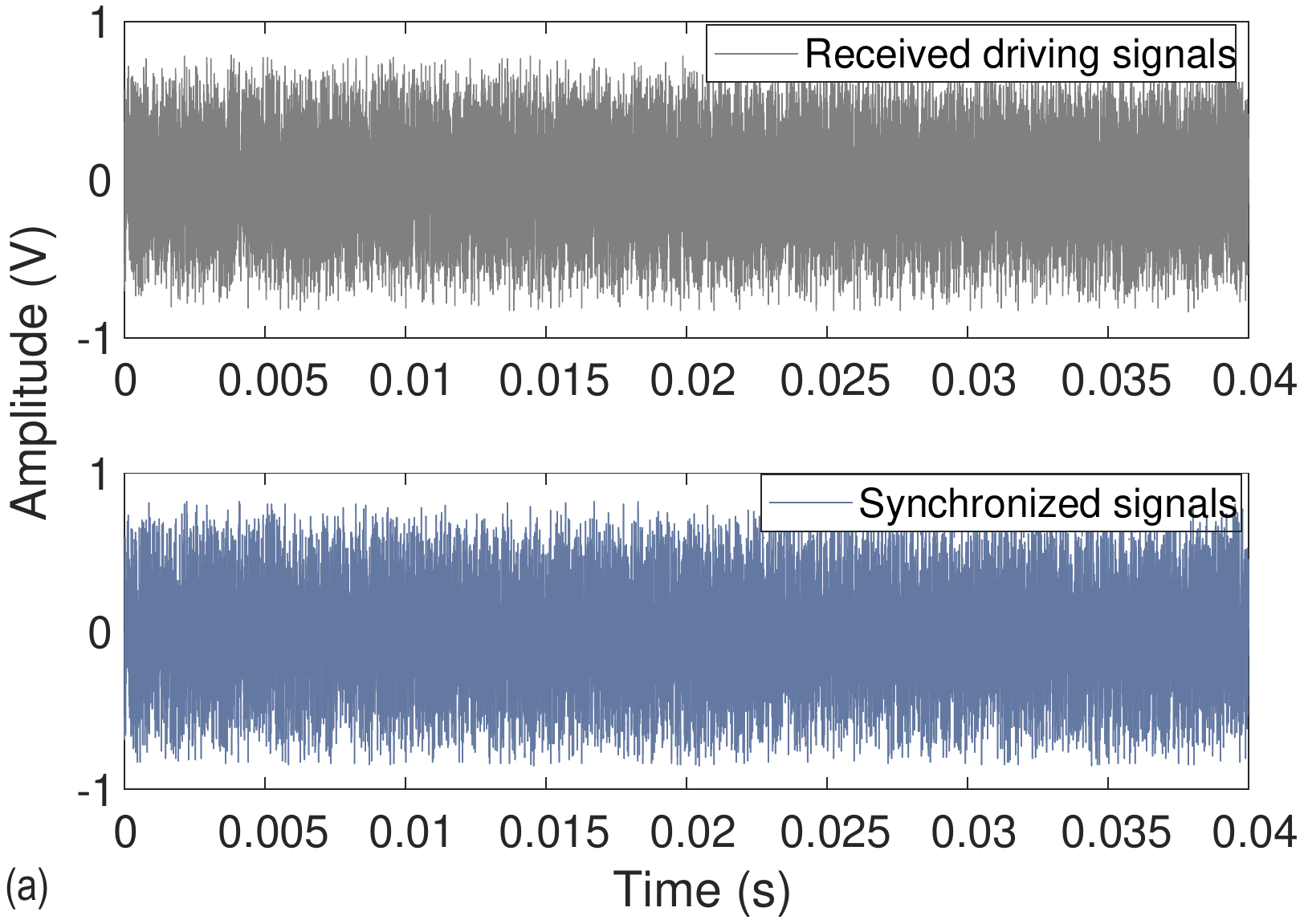}~~~~~
     \includegraphics[width=0.44\textwidth]{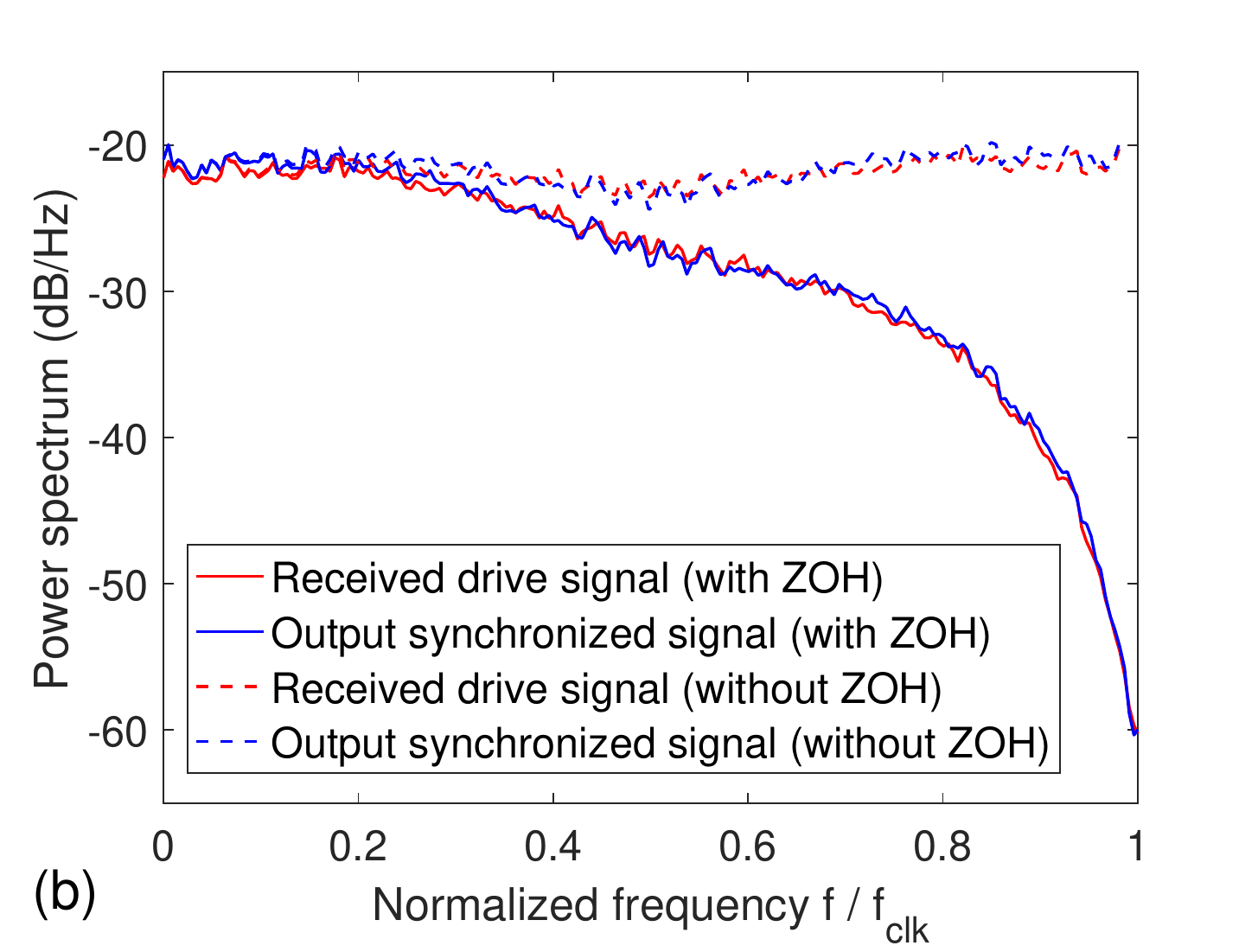}
	\end{center}
	\caption{(a) Transient signals for the scalar drive signal $w^{*t}_{n}$ and its synchronized signal on receiver side $w^{r}_{n}$, where `$*$' denotes the variable mixed with information. $f_{clk}=0.5$~MHz is implemented. (b) Their corresponding power spectral densities with and without the output zero-order hold (ZOH) estimated using Welch's method.}
	\label{fig:transient_signal}
\end{figure}

Fig.~\ref{fig:transient_signal}(a) shows a typical transmitted drive signal $w^{*t}_{n}$ and its synchronized response signal $w^{r}_{n}$ in the time domain. Because of the VP or VE properties of the designed circuit system when using Eqn.~(\ref{system_eqn}), no specific structures are present and the drive and response signals ($w^{*t}_{n}$ and $w^{*r}_{n}$, respectively) behave like white noise. Thus, their corresponding frequency spectra, as shown in Fig.~\ref{fig:transient_signal}(b), overlap with each other and do not fall off until $f_{clk}/2$. Note that the observed spectra differ from the simulated ones shown in Fig.~\ref{fig:hyperchaos_map_2_phase_psd}(b) because the former are continuous-time signals that include a zero-order hold (ZOH) between switching instants. Thus, the observed spectra are filtered by the frequency response of the ZOH, which is given by
\begin{equation}
H_{ZOH}(f)=e^{-j\pi f/f_{clk}}\frac{sin\left(\pi f/f_{clk}\right)}{\pi f/f_{clk}}.
\end{equation} 
It is clear that $H_{ZOH}(f)$ is a low-pass filter with nulls at integer multiples of $f_{clk}$, which explains the observed high-frequency roll-off in Fig.~\ref{fig:transient_signal}(b). Its effects can be removed by dividing the observed power spectra by $\left|H_{ZOH}(f)\right|^{2}$, as shown in Fig.~\ref{fig:transient_signal}(b). The resulting spectra are almost white, as expected from simulations.

The observed properties of the proposed synchronized maps, including: i) little structure or pattern in the attractor; ii) short correlation time and fast synchronization (both within a few clock periods); iii) sensitivity to circuit parameter mismatch; and iv) hyperchaotic dynamics, enhance the privacy of the digital communication system via chaotic masking and pose extreme difficulty in decoding without using a parameter-matched receiver. For example, small mismatches in the coefficients of the transformation matrix $\mathbf{A}$ in Eqn. (\ref{system_eqn}), which can be created simply by tuning resistor values in the addition or subtraction circuits, will lead to de-synchronization between the transmitter and receiver.

\begin{figure}[h]
	\begin{center}
		\includegraphics[width=0.46\textwidth]{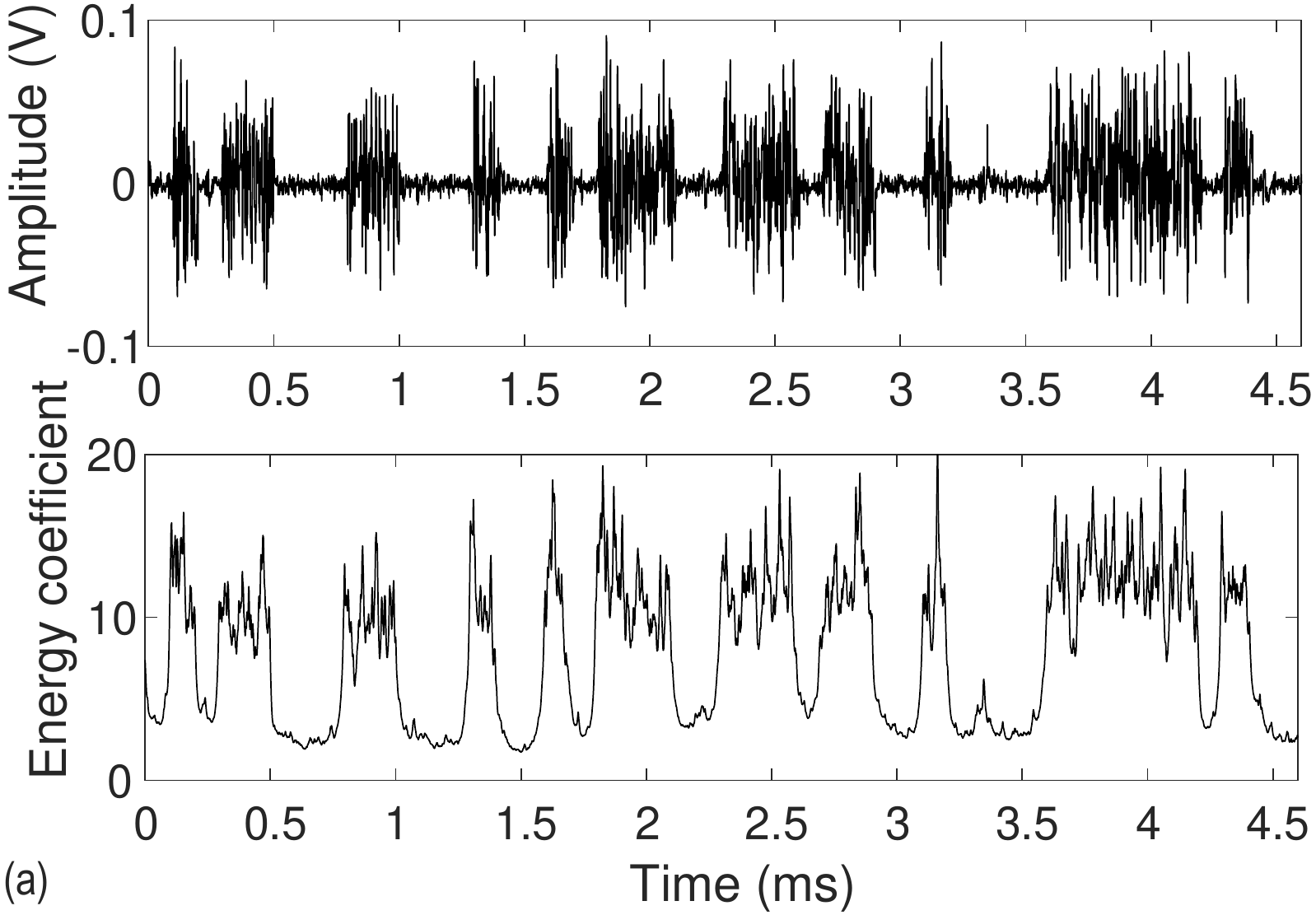}~~~~~
		\includegraphics[width=0.46\textwidth]{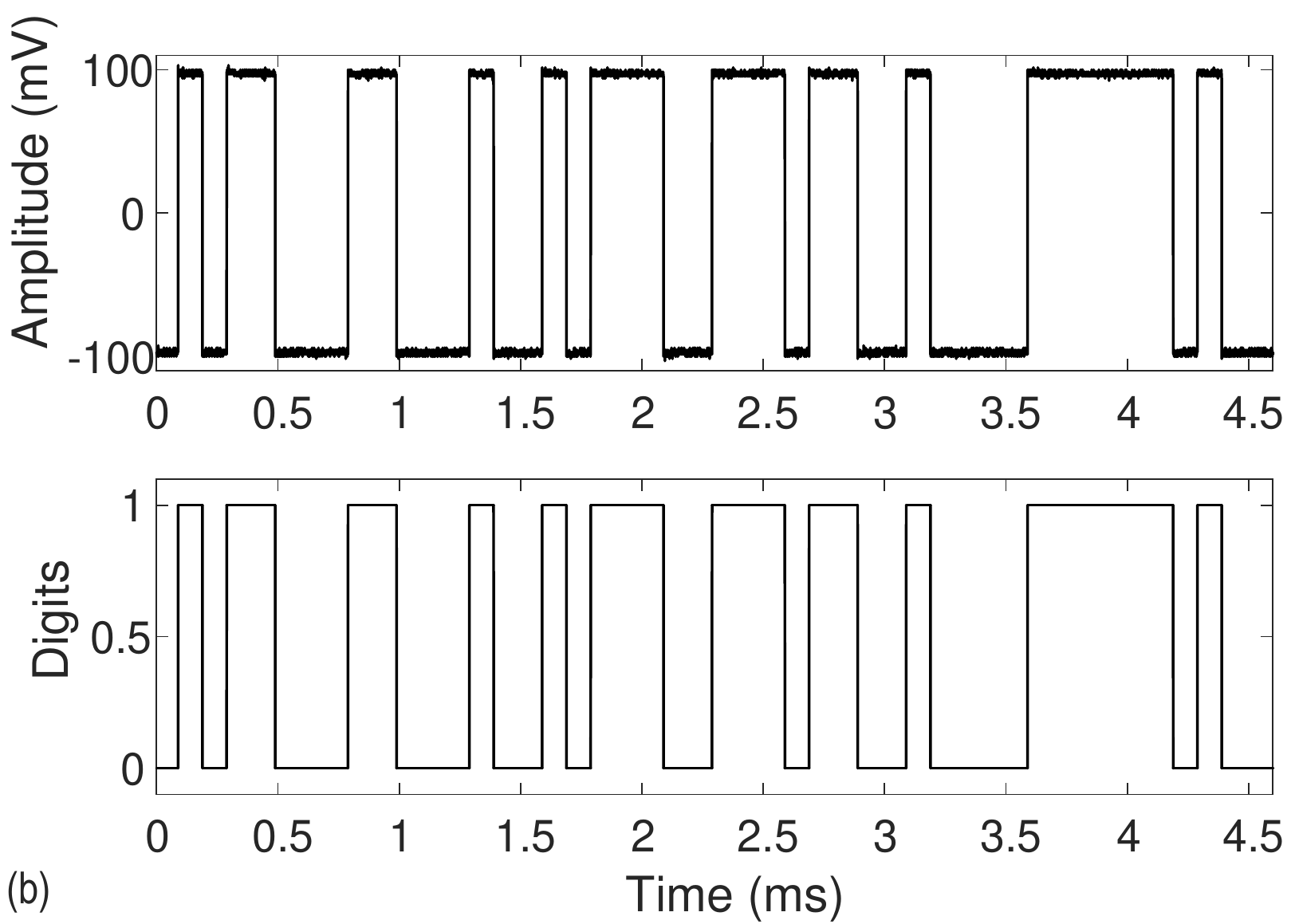}
	\end{center}
	\caption{(a) Demodulated digital signals through $w^{*t}_{n}-w^{r}_{n}$ (top) when systems are in synchronization. Envelope extraction for digital signal recovery using CWT with Morlet wavelet. (b) Digital signals (top) to be mixed with the transmitter variable $z^{t}_{n}$ and retrieved digital information through adaptive thresholding on enveloped signals in (a) after matched filtering (an integrate and dump receiver).}
	\label{fig:digital_data}
\end{figure}

Fig.~\ref{fig:digital_data} shows the process of demodulation/unmasking on the received chaotic masking signals when mixed with a digital signal amplitude of 200~mV for a transmission bit rate of 10~kbps and $f_{clk}=0.5$~MHz. To extract the digital information, we firstly unmask the digital signal using the synchronized signals of the receiver, as shown in Fig.~\ref{fig:digital_data}(a) (top) and then implement a continuous wavelet transform (CWT) to extract the envelope of the digital information, as shown in Fig.~\ref{fig:digital_data}(a) (bottom). Bits can be further obtained through thresholding of the envelope. Fig.~\ref{fig:digital_data} (b) shows the extracted digital information (bottom) after thresholding of the filtered envelope. Good agreement between transmitted and recovered information as illustrated in Fig.~\ref{fig:digital_data}(b), verifying the effectiveness of our proposed digital chaotic communication system.

We generate pseudorandom binary sequences ($1$~Mbits, 200~mV and $f_{clk}=0.5$~MHz) as test data streams that are masked with chaotic signals to quantitatively study adaptive energy-based thresholding for the recovery process shown in Fig.~\ref{fig:digital_data}(a). The well-known integrate and dump filter (IDF) is implemented in this paper. The IDF is used as a matched filter for coherent detection of rectangular pulse shapes (symbols) corrupted by additive white Gaussian noise (AWGN) when symbol transition times are given (i.e., the transmitter and receiver have synchronous clocks). It can be replaced by the appropriate matched filter if other symbol shapes are used.

\begin{figure}[h]
	\begin{center}
		\includegraphics[width=0.33\textwidth]{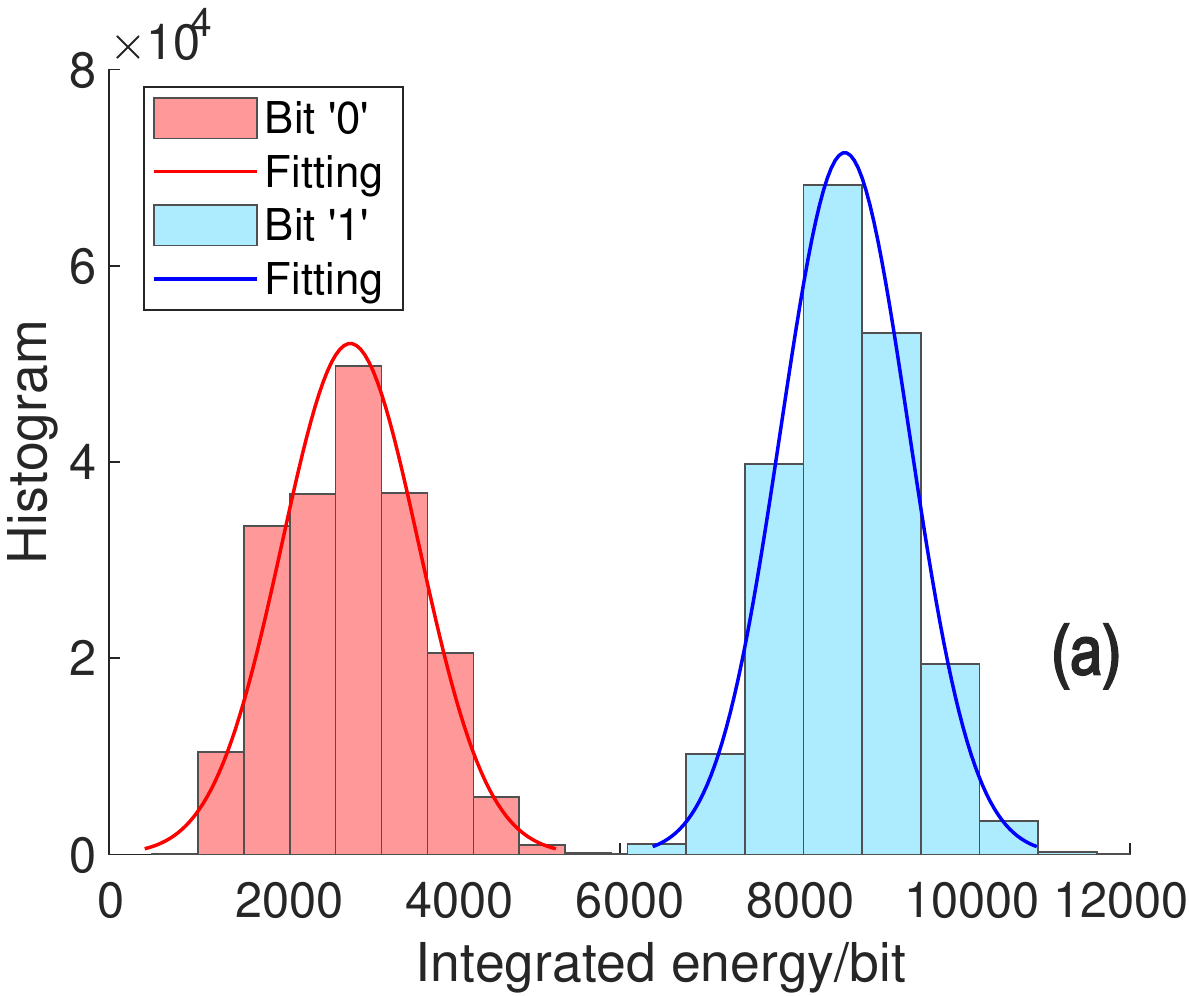}
		\includegraphics[width=0.33\textwidth]{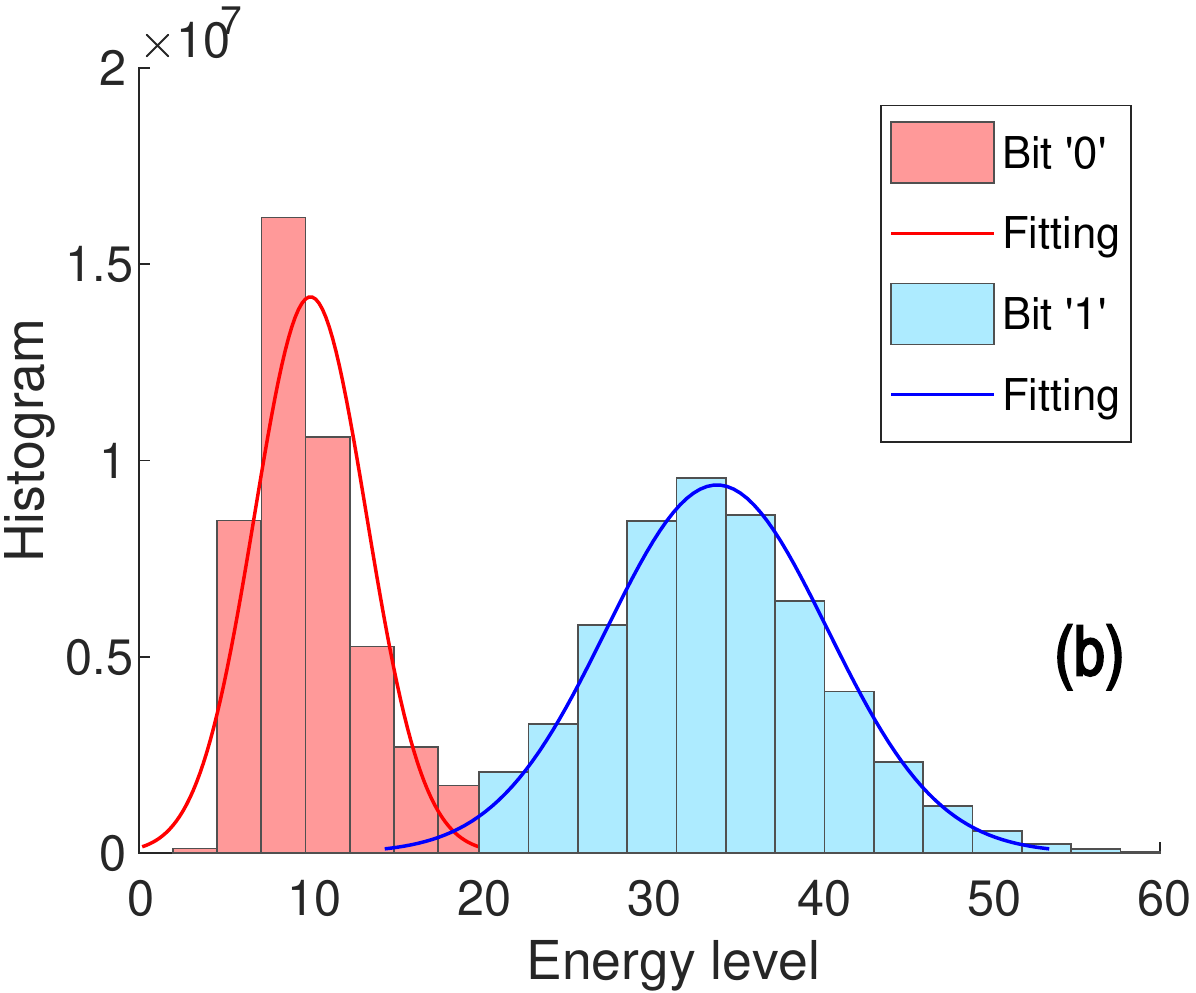}
		\includegraphics[width=0.32\textwidth]{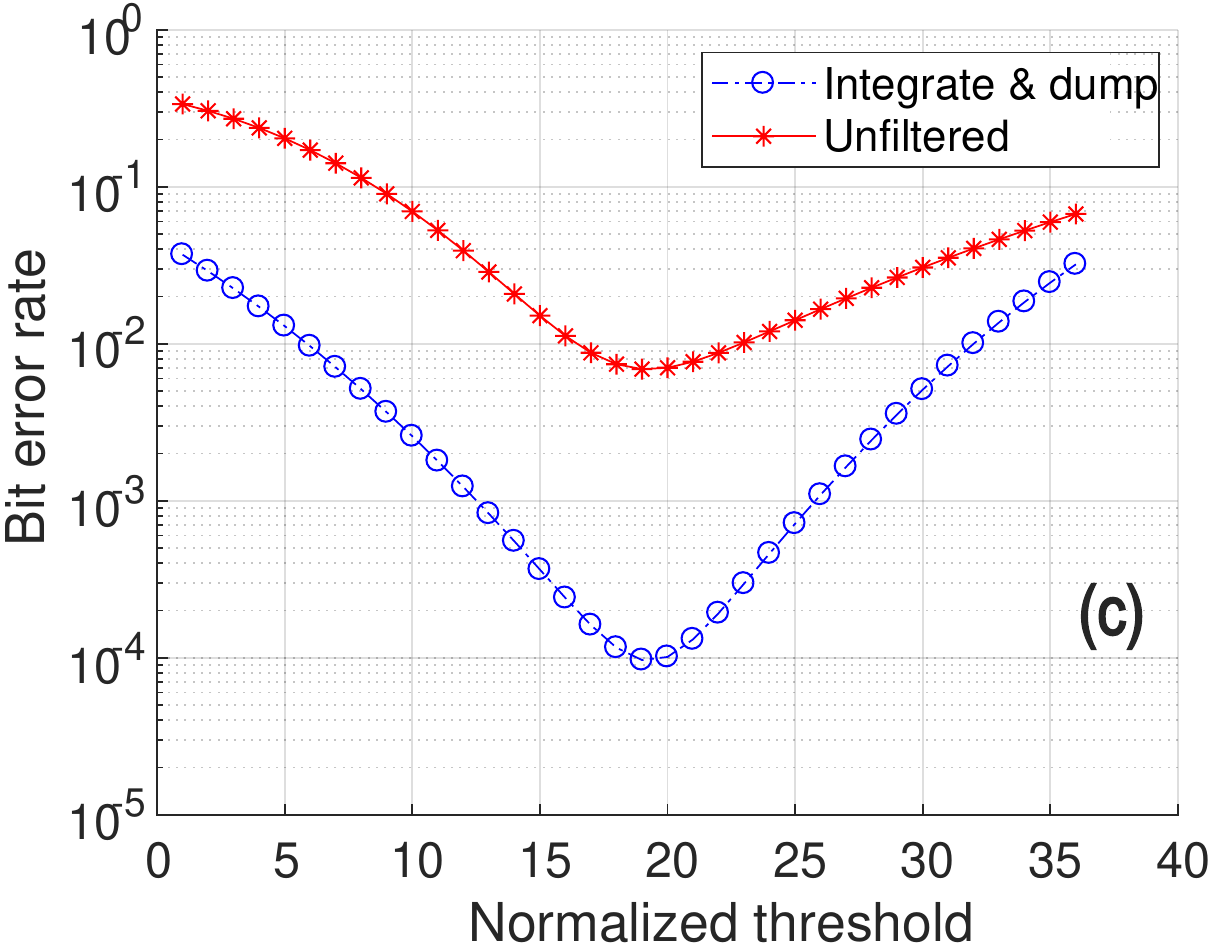}
	\end{center}
	\caption{Adaptive energy-based thresholding based on an integrate-and-dump filter for bit stream reconstruction when $f_{clk}=0.5$~MHz and $R_{b}=10$~kbps: (a) histogram for bits `0' and `1' after integrate-and-dump filtering; (b) histogram for bits `0' and `1' without any filtering; (c) bit error rate (BER) estimation for each scenario when using different thresholds.}
	\label{fig:data_thre}
\end{figure}

Each symbol occupies $N=f_{clk}/R_s$ clock periods where $R_s$ is the symbol rate; note that $R_{s}$ is equal to the bit rate $R_{b}$ in this case since we transmit binary symbols. Figs.~\ref{fig:data_thre}(a) and (b) show typical measured histograms for symbols `0' (red) and `1' (blue) with and without matched filtering, respectively. Both histograms are approximately Gaussian when $N\gg 1$ (as in this case, for which $N=50$) due to the central limit theorem. This suggests that an IDF will act as a nearly-ideal receiver for the proposed channel when $R_{b}\ll f_{clk}$.

Considering a non-return to zero (NRZ) transmission scheme with Gaussian measurement noise of $\omega_{i}\sim\mathcal{N}(0,\sigma_{i}),i=0,1$, we have $x_{0}\sim\mathcal{N}(\mu_{0},\sigma_{0})$ for symbol `0' and $x_{1}\sim\mathcal{N}(\mu_{1},\sigma_{1})$ for symbol `1', where $\mu_{0,1}$ are the estimated mean signal amplitudes for symbols `0' and `1', respectively. Therefore, the likelihood of a bit misinterpretation can be expressed as follows:
\begin{equation}
p_{BER} = p(0|1)p(1)+p(1|0)p(0),
\label{ber}
\end{equation}
where $p(0) = N_{0}/(N_{0}+N_{1})$ is the prior probability for bit `0' and $p(1) = 1-p(0)$ is that for bit `1'; here $N_{i},i=0,1$ are the number of transmitted bits for $i=0,1$ during experiments. The conditional error probabilities are written as
\begin{eqnarray}
p(1|0) = \frac{1}{2}\mathrm{\bf{erfc}}\left(\frac{\lambda-\mu_{0}}{\sigma_{0}\sqrt{2}}\right)\\
p(0|1) = \frac{1}{2}\mathrm{\bf{erfc}}\left(\frac{-\lambda+\mu_{1}}{\sigma_{1}\sqrt{2}}\right),
\label{eq:ber2}
\end{eqnarray}
where $\mathrm{\bf{erfc}}$ is the complementary error function and $\lambda$ is the selected energy threshold. Fig.~\ref{fig:data_thre}(c) illustrates the bit error rate (BER) predicted by Eqn.~(\ref{ber}) for both filtered and unfiltered waveforms while varying the threshold level. The parameters $\mu$ and $\sigma$ are estimated from the fitted Gaussian distributions shown in Figs.~\ref{fig:data_thre}(a) and (b). The integrate-and-dump filtered receiver ($\mathrm{BER_{opt}}$ = $10^{-4}$) significantly outperforms the unfiltered receiver ($\mathrm{BER_{opt}}$ = $10^{-2}$), i.e., matched filtering improves BER by $100\times$ in this case. This may also be explained as the result of averaging: the IDF receiver effectively averages the received signal over $N\gg 1$ clock periods before thresholding. This improves the effective signal-to-noise ratio (SNR) for making bit decisions (i.e., the arguments of the error functions in Eqn.~(\ref{eq:ber2})) by $\sqrt{N}$ if we assume that the hyperchaotic transmitter generates i.i.d. random variables. The latter is evidently a reasonable approximation. Adaptability can be achieved through iteratively updating the optimal threshold based on accumulated experimental bit streams, where the noise level is learned along the time series.

\begin{figure}[h]
	\begin{center}
		\includegraphics[width=0.43\textwidth]{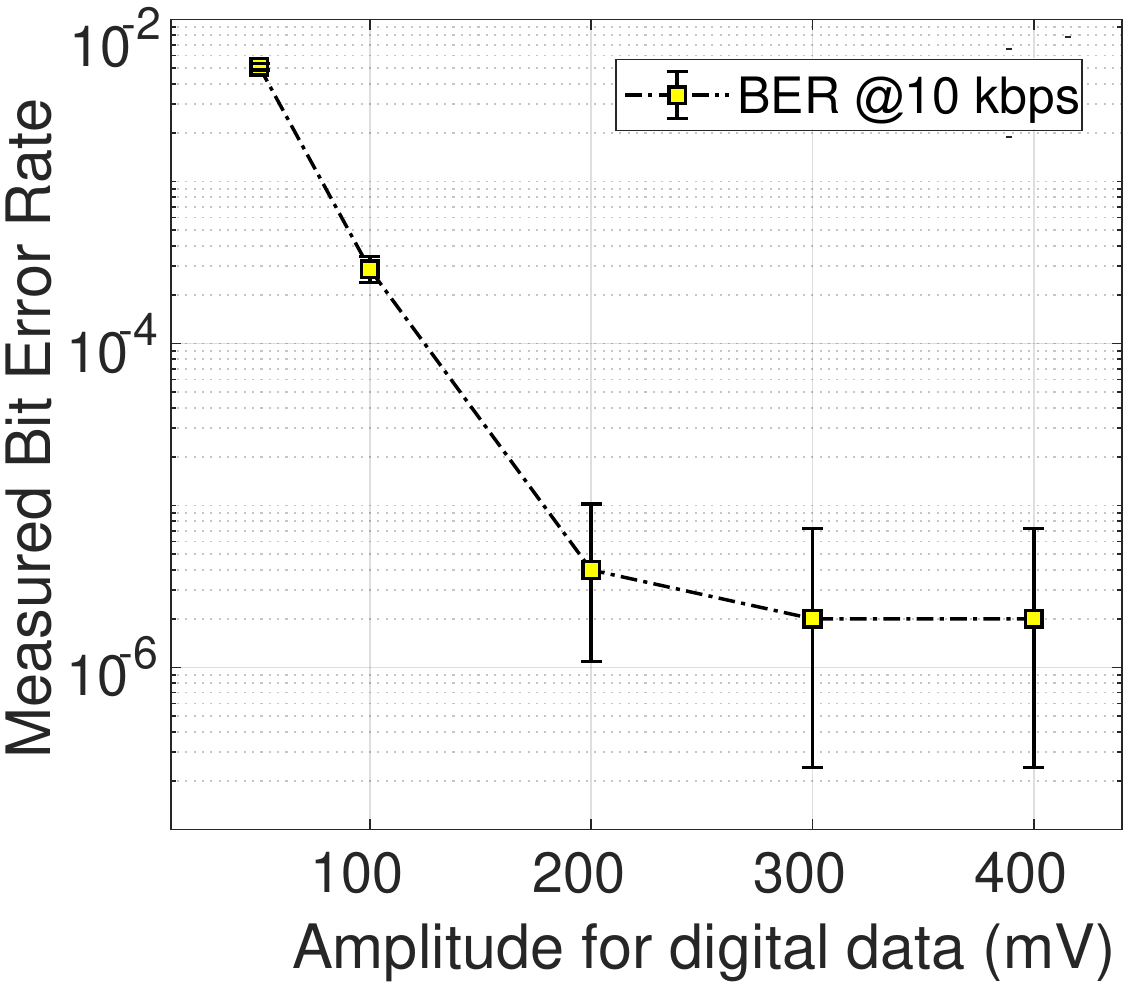}
	\end{center}
	\caption{Measured BER for the proposed digital hyperchaotic masking communication link after matched filtering as a function of the amplitude of the transmitted digital pulses. The clock frequency $f_{clk}=0.5$~MHz and the transmission bit rate $R_{b}=10$~kbps.}
	\label{fig:data_ber}
\end{figure}

Fig.~\ref{fig:data_ber} summarizes the measured BER of the proposed digital communication link after matched filtering as a function of the transmitted signal amplitude $d_{in}$. The BER decreases exponentially with $d_{in}$ as predicted by Eqn.~(\ref{eq:ber2}) before appearing to saturate at a residual level of $\approx 2\times 10^{-6}$, where 95$\%$ confidence intervals are estimated from finite sets of test bits (1~Mbits). Note that the limited sizes of these sets result in significant BER estimation errors for signal amplitudes $>200$~mV, i.e., not enough bit errors occur to allow accurate estimation of BER. Thus, the observed minimum BER of $\approx 2\times 10^{-6}$ is probably limited by the finite lengths of the test bit streams. 
 
\begin{figure}[h]
	\begin{center}
		\includegraphics[width=0.45\textwidth]{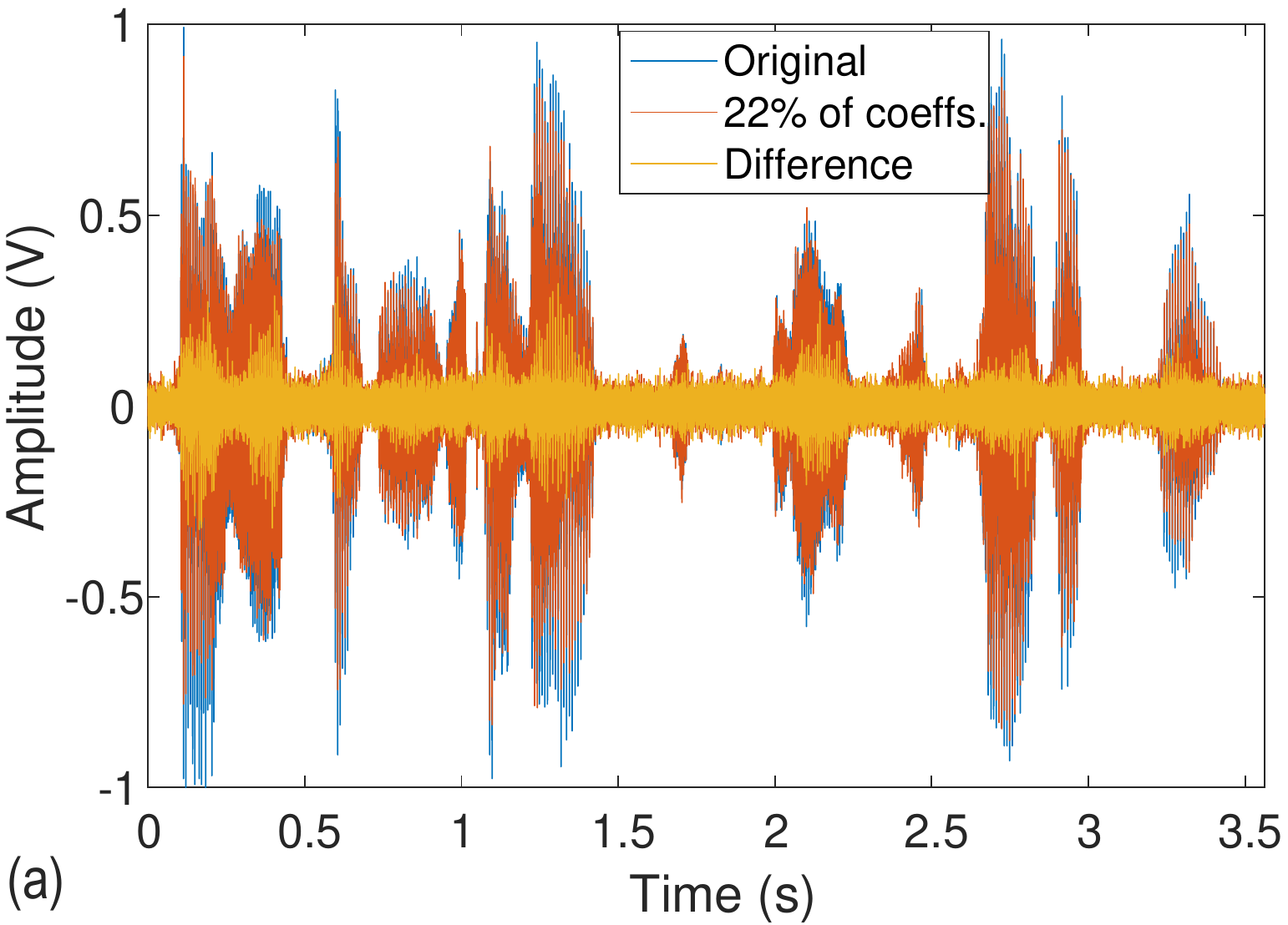}~
		\includegraphics[width=0.50\textwidth]{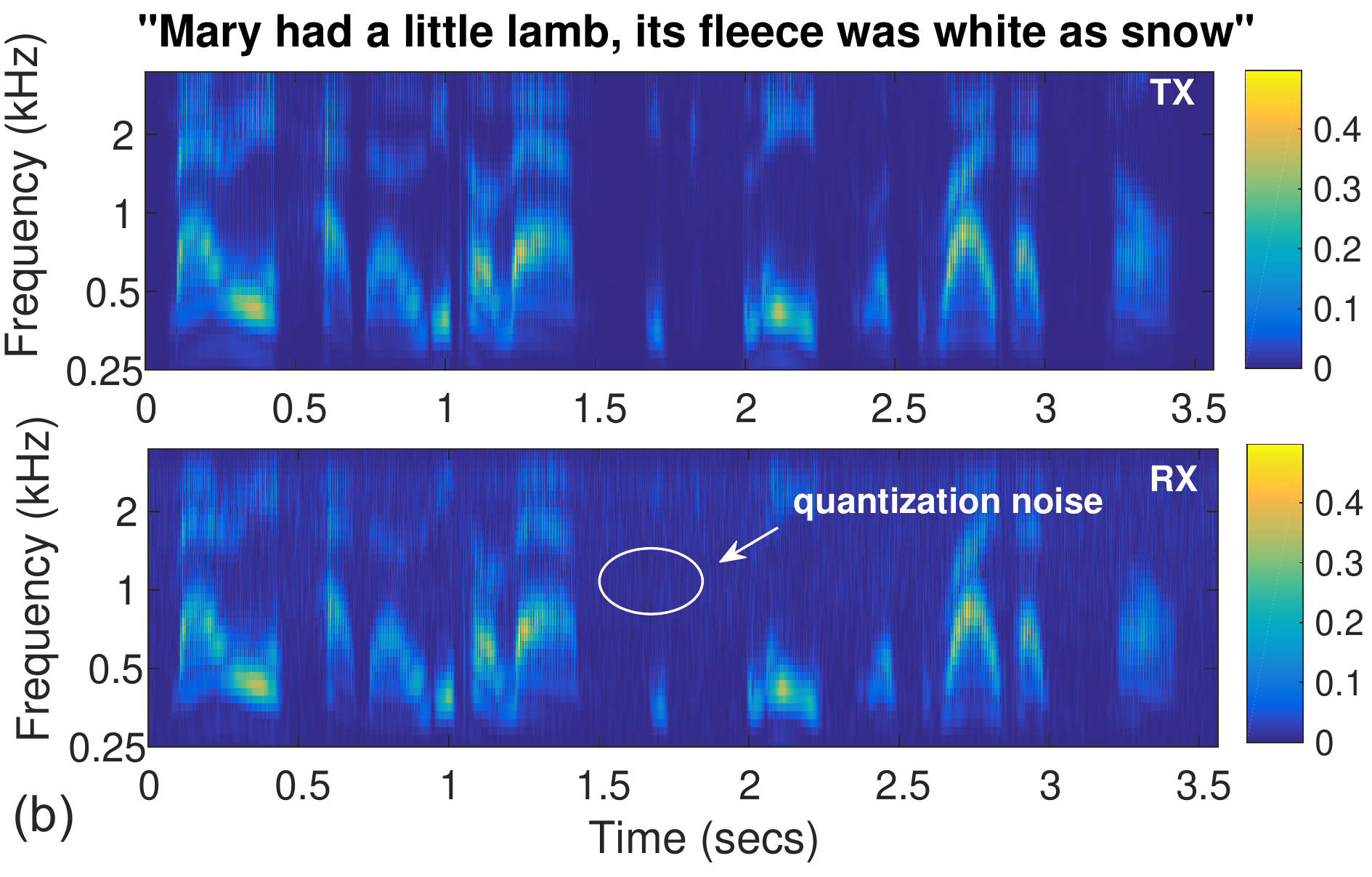}
	\end{center}
	\caption{(a) Recorded audio signals using the microphone on a PC at a sampling rate $f_{s}=8$~kHz and 12~bits per sample: original audio, DCT-compressed signals after reconstruction, and their difference; (b) frequency-time domain spectrograms of the transmitted and recovered audio signals using CWT.}
	\label{fig:speech_case}
\end{figure}

We further examine the functionality of our proposed digital chaotic-masking/unmasking communication system using two case studies: (i) real-time speech transmission; (ii) gray-scale image transmission. In both experiments, we implement a data bit rate of 14~kbps, $f_{clk}=0.5$~MHz, and a digital signal amplitude of 200~mV. In the speech case, we used a common microphone (12-bit ADC, sampling rate of 8~kHz, i.e., 96~kbps) on a PC to record a volunteer speaking the sentence "Mary had a little lamb, its fleece was white as snow". To achieve real-time transmission given limited channel bandwidth and data rate, a discrete cosine transform (DCT) was implemented to compress the down-converted (8-bit) recorded speech, where only 22\% of the original coefficients were used. The resulting compression ratio (CR) was up to 4.55, as shown in Fig.~\ref{fig:speech_case} (a). This compression scheme reduces data rate down to 14.1~kbps, which is low enough to be transmitted in real-time through the proposed link. Small differences (less than 3\%) cased by DCT compression can be ignored. Fig.~\ref{fig:speech_case} (b) shows the frequency-time domain spectrogram before compression and after chaotic unmasking and recovery, respectively.  Although a small amount of quantization noise can be observed in spectrum, legible speech can be clearly heard when replaying the retrieved chaotic unmasked signals after the inverse DCT.

\begin{figure}[h]
	\begin{center}
		\includegraphics[width=0.38\textwidth]{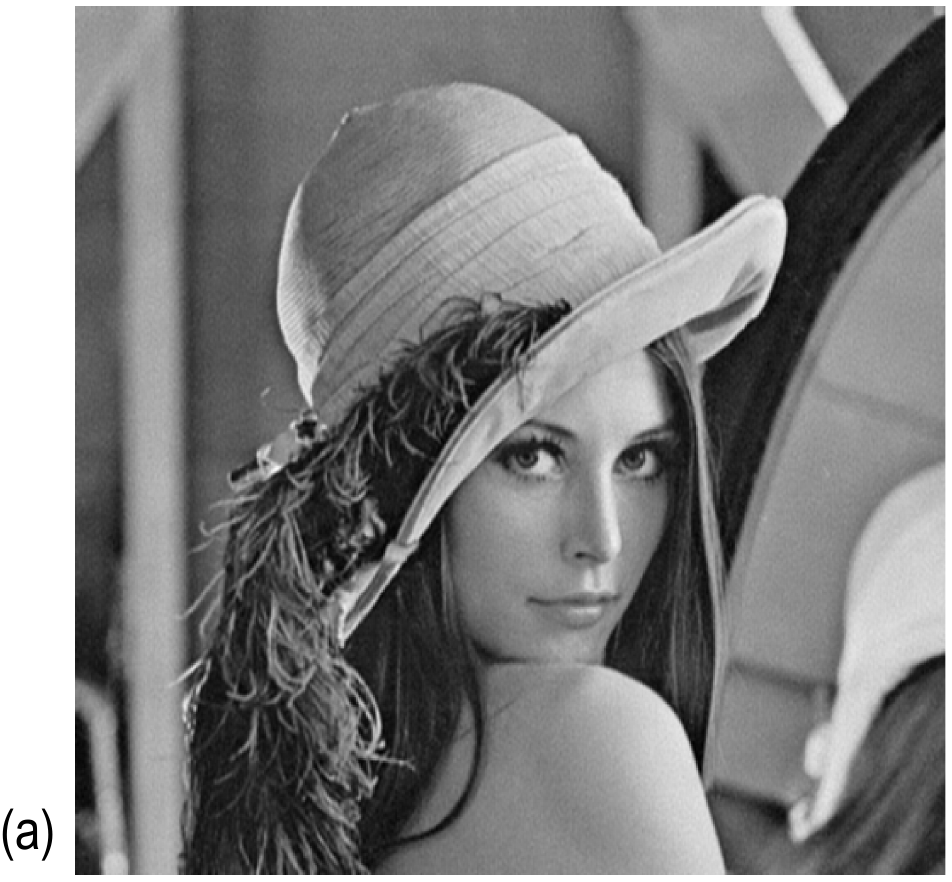}~~~~~~~~
		\includegraphics[width=0.38\textwidth]{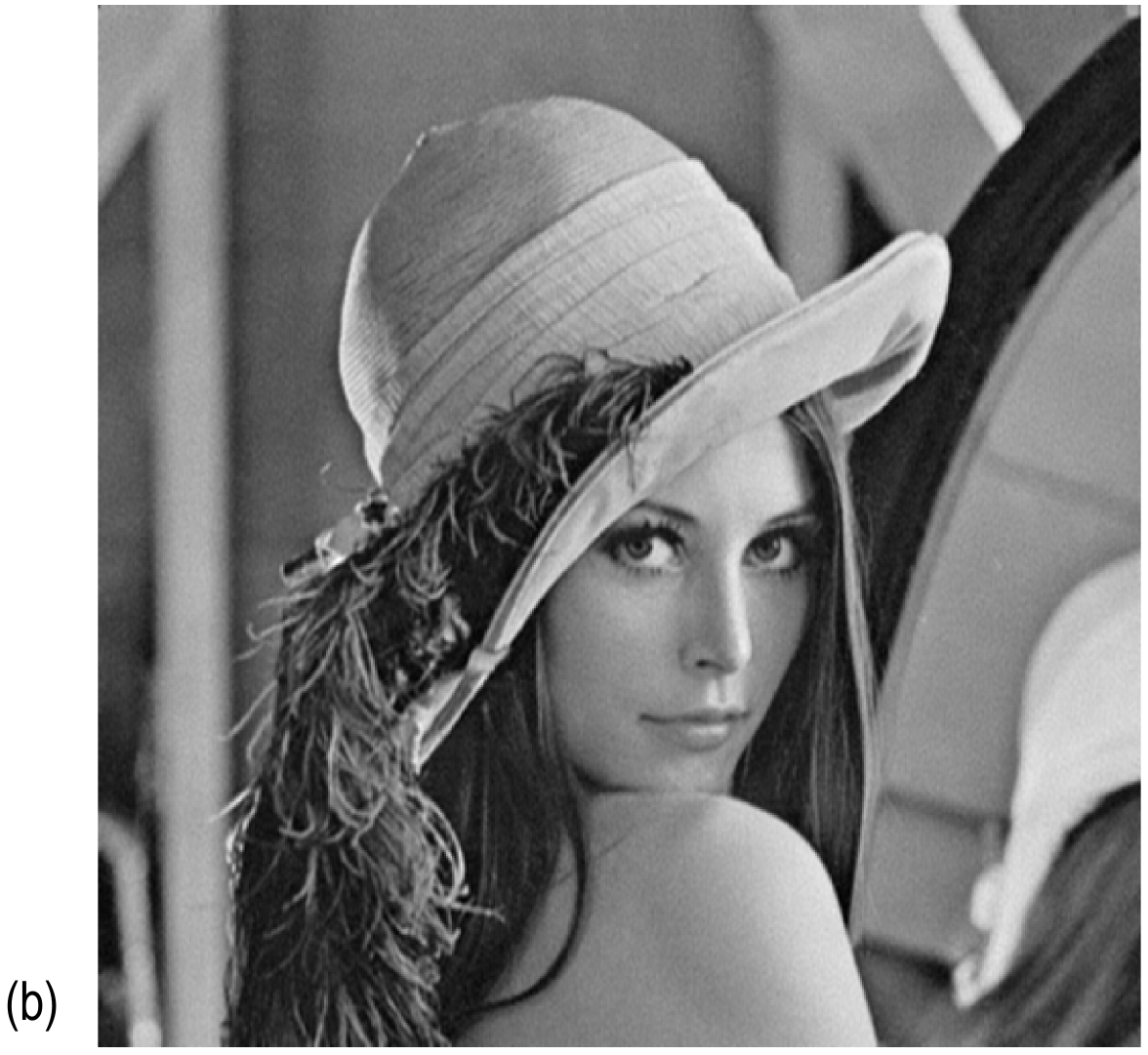}	
	\end{center}
	\caption{(a) 2D-DCT compressed image used for testing the proposed private digital communication link using synchronized hyperchaotic systems. Only 16.5~\% of the coefficients of the original image were used for chaotic-masking transmission. (b) Recovered image using inverse 2D-DCT on the retrieved coefficients.}
	\label{fig:image_case}
\end{figure}

In the image scenario, the 2D-DCT is used to compressed the original images. For example, the common image shown in Fig.~\ref{fig:image_case}(a) resulted in a CR of 6.1, which significantly reduces the transmission load. Fig.~\ref{fig:image_case} (b) illustrates the recovered gray-scale image, which is essentially identical to the original one. Successful transmission with chaotic masking/unmasking on speech and image cases further verifies the effectiveness, robustness, and reliability of our proposed digital communication scheme using synchronized hyperchaotic systems.

\section{Conclusion}
\label{sec:conclusion}
We have described a private digital communication system based on synchronized hyperchaotic maps. Two such maps were implemented at the board level and tested at clock frequencies up to 1.5~MHz. The synchronized maps were used for communication via chaotic masking and matched filtering. Excellent BER of $\approx 2\times 10^{-6}$ was obtained at bit rates up to 14~kbps, allowing real-time transmission of compressed speech without further error correction. Gray scale images were also successfully transmitted. The maximum clock frequency can be further increased (up to at least $\sim$5~MHz) by reducing the inter-phase and sample-and-hold delay times on the boards, thus allowing data rates up to $\sim$100~kbps.  

In future work, we will consider custom integrated circuit (IC) implementations of the proposed hyperchaotic system. Such single-chip designs will enable further increases in clock speed and data rates while reducing system size and power consumption.

\nonumsection{Acknowledgments} 
\noindent The authors wish to thank Cheng Chen and David Ariando for assistance with board ordering and assembly.


\bibliographystyle{ws-ijbc}
\bibliography{ijbc_2017}

\begin{thebibliography}{28}
\newcommand{\enquote}[1]{``#1''}
\providecommand{\natexlab}[1]{#1}
\providecommand{\url}[1]{\texttt{#1}}
\providecommand{\urlprefix}{URL }
\expandafter\ifx\csname urlstyle\endcsname\relax
  \providecommand{\doi}[1]{doi:\discretionary{}{}{}#1}\else
  \providecommand{\doi}{doi:\discretionary{}{}{}\begingroup
  \urlstyle{rm}\Url}\fi

\bibitem[{Brown \emph{et~al.}(1994)Brown, Rulkov \& Tufillaro}]{Brown1994}
Brown, R., Rulkov, N.~F. \& Tufillaro, N.~B. [1994] \enquote{Synchronization of
  chaotic systems: The effects of additive noise and drift in the dynamics of
  the driving,} \emph{Phys. Rev. E} \textbf{50},  4488--4508,
  \doi{10.1103/PhysRevE.50.4488}.

\bibitem[{Carroll \emph{et~al.}(1996)Carroll, Heagy \& Pecora}]{Carroll1996}
Carroll, T.~L., Heagy, J.~F. \& Pecora, L.~M. [1996] \enquote{Transforming
  signals with chaotic synchronization,} \emph{Phys. Rev. E} \textbf{54},
  4676--4680, \doi{10.1103/PhysRevE.54.4676}.

\bibitem[{Carroll \& Pecora(1998)}]{Carroll1998}
Carroll, T.~L. \& Pecora, L.~M. [1998] \enquote{Synchronizing hyperchaotic
  volume-preserving maps and circuits,} \emph{IEEE Transactions on Circuits and
  Systems I: Fundamental Theory and Applications} \textbf{45},  656--659,
  \doi{10.1109/81.678482}.

\bibitem[{Chen(2003)}]{Chen2003}
Chen, G. [2003]  \emph{Chaotification via Feedback: The Discrete Case}
  (Springer Berlin Heidelberg, Berlin, Heidelberg), ISBN 978-3-540-44986-7, pp.
  159--177, \doi{10.1007/978-3-540-44986-7_8}.

\bibitem[{Chen \& Shi(2006)}]{Chen2006}
Chen, G. \& Shi, Y. [2006] \enquote{Introduction to anti-control of discrete
  chaos: theory and applications,} \emph{Philosophical Transactions of the
  Royal Society of London A: Mathematical, Physical and Engineering Sciences}
  \textbf{364},  2433--2447, \doi{10.1098/rsta.2006.1833}.

\bibitem[{Chen \emph{et~al.}(2011)Chen, Liang \& Tong}]{Chen2011}
Chen, Z.-Q., Liang, G.-Q. \& Tong, J.-G. [2011] \enquote{The {FPGA}
  implementation of hyperchaotic system based upon {VHDL} design,}  \emph{2011
  Fourth International Workshop on Chaos-Fractals Theories and Applications},
  pp. 47--51, \doi{10.1109/IWCFTA.2011.82}.

\bibitem[{Eckmann \emph{et~al.}(1986)Eckmann, Kamphorst, Ruelle \&
  Ciliberto}]{Eckmann1986}
Eckmann, J.~P., Kamphorst, S.~O., Ruelle, D. \& Ciliberto, S. [1986]
  \enquote{Liapunov exponents from time series,} \emph{Phys. Rev. A}
  \textbf{34},  4971--4979, \doi{10.1103/PhysRevA.34.4971}.

\bibitem[{Elwakil \& Kennedy(1999)}]{Elwakil1999}
Elwakil, A. \& Kennedy, M. [1999] \enquote{Inductorless hyperchaos generator,}
  \emph{Microelectronics Journal} \textbf{30},  739 -- 743,
  \doi{https://doi.org/10.1016/S0026-2692(98)00163-3}.

\bibitem[{Filali \emph{et~al.}(2014)Filali, Benrejeb \& Borne}]{Filali2014}
Filali, R.~L., Benrejeb, M. \& Borne, P. [2014] \enquote{On observer-based
  secure communication design using discrete-time hyperchaotic systems,}
  \emph{Communications in Nonlinear Science and Numerical Simulation}
  \textbf{19},  1424--1432.

\bibitem[{Grassberger \& Procaccia(1983)}]{Grassberger1983}
Grassberger, P. \& Procaccia, I. [1983] \enquote{Characterization of strange
  attractors,} \emph{Phys. Rev. Lett.} \textbf{50},  346--349,
  \doi{10.1103/PhysRevLett.50.346}.

\bibitem[{Grassi \emph{et~al.}(2012)Grassi, Cafagna, Vecchio \&
  Miller}]{Grassi2012}
Grassi, G., Cafagna, D., Vecchio, P. \& Miller, D.~A. [2012] \enquote{A new
  scheme to synchronize chaotic discrete-time systems via a scalar signal,}
  \emph{Circuits and Systems (MWSCAS), 2012 IEEE 55th International Midwest
  Symposium on} (IEEE), pp. 654--657.

\bibitem[{Grassi \& Mascolo(1999)}]{Grassi1999}
Grassi, G. \& Mascolo, S. [1999] \enquote{A system theory approach for
  designing cryptosystems based on hyperchaos,} \emph{IEEE Transactions on
  Circuits and Systems I: Fundamental Theory and Applications} \textbf{46},
  1135--1138.

\bibitem[{Grassi \& Miller(2002)}]{Grassi2002}
Grassi, G. \& Miller, D.~A. [2002] \enquote{Theory and experimental realization
  of observer-based discrete-time hyperchaos synchronization,} \emph{IEEE
  Transactions on Circuits and Systems I: Fundamental Theory and Applications}
  \textbf{49},  373--378, \doi{10.1109/81.989174}.

\bibitem[{Hassan(2014)}]{Hassan2014}
Hassan, M.~F. [2014] \enquote{A new approach for secure communication using
  constrained hyperchaotic systems,} \emph{Applied Mathematics and Computation}
  \textbf{246},  711--730.

\bibitem[{Kolumban \emph{et~al.}(1996)Kolumban, Vizv{\'a}ri, Schwarz \&
  Abel}]{Kolumban1996}
Kolumban, G., Vizv{\'a}ri, B., Schwarz, W. \& Abel, A. [1996]
  \enquote{Differential chaos shift keying: {A} robust coding for chaos
  communication,}  \emph{Proc. NDES}, pp. 87--92.

\bibitem[{Matsumoto \emph{et~al.}(1986)Matsumoto, Chua \&
  Kobayashi}]{Matsumoto1986}
Matsumoto, T., Chua, L. \& Kobayashi, K. [1986] \enquote{Hyper chaos:
  {L}aboratory experiment and numerical confirmation,} \emph{IEEE Transactions
  on Circuits and Systems} \textbf{33},  1143--1147,
  \doi{10.1109/TCS.1986.1085862}.

\bibitem[{Oseledets(1968)}]{Oseledets1968}
Oseledets, V.~I. [1968] \enquote{A multiplicative ergodic theorem.
  {C}haracteristic {L}yapunov exponents of dynamical systems,} \emph{Trudy
  Moskovskogo Matematicheskogo Obshchestva (Transactions of the Moscow
  Mathematical Society)} \textbf{19},  179--210.

\bibitem[{Pecora \emph{et~al.}(1997)Pecora, Carroll, Johnson \&
  Mar}]{Pecora1997}
Pecora, L.~M., Carroll, T.~L., Johnson, G. \& Mar, D. [1997]
  \enquote{Volume-preserving and volume-expanding synchronized chaotic
  systems,} \emph{Phys. Rev. E} \textbf{56},  5090--5100,
  \doi{10.1103/PhysRevE.56.5090}.

\bibitem[{Peng \emph{et~al.}(1996)Peng, Ding, Ding \& Yang}]{Peng1996}
Peng, J.~H., Ding, E.~J., Ding, M. \& Yang, W. [1996] \enquote{Synchronizing
  hyperchaos with a scalar transmitted signal,} \emph{Phys. Rev. Lett.}
  \textbf{76},  904--907, \doi{10.1103/PhysRevLett.76.904}.

\bibitem[{R{\"o}ssler(1979)}]{Rossler1979a}
R{\"o}ssler, O. [1979] \enquote{Chaotic oscillations: {A}n example of
  hyperchaos,} \emph{Nonlinear Oscillations in Biology} \textbf{17},  141--156.

\bibitem[{Sano \& Sawada(1985)}]{Sano1985}
Sano, M. \& Sawada, Y. [1985] \enquote{Measurement of the {L}yapunov spectrum
  from a chaotic time series,} \emph{Phys. Rev. Lett.} \textbf{55},
  1082--1085, \doi{10.1103/PhysRevLett.55.1082}.

\bibitem[{Shen \emph{et~al.}(2014{\natexlab{a}})Shen, Yu, Lü \&
  Chen}]{Shen2014b}
Shen, C., Yu, S., Lü, J. \& Chen, G. [2014{\natexlab{a}}] \enquote{Designing
  hyperchaotic systems with any desired number of positive lyapunov exponents
  via a simple model,} \emph{IEEE Transactions on Circuits and Systems I:
  Regular Papers} \textbf{61},  2380--2389, \doi{10.1109/TCSI.2014.2304655}.

\bibitem[{Shen \emph{et~al.}(2014{\natexlab{b}})Shen, Yu, Lü \&
  Chen}]{Shen2014}
Shen, C., Yu, S., Lü, J. \& Chen, G. [2014{\natexlab{b}}] \enquote{A
  systematic methodology for constructing hyperchaotic systems with multiple
  positive {L}yapunov exponents and circuit implementation,} \emph{IEEE
  Transactions on Circuits and Systems I: Regular Papers} \textbf{61},
  854--864, \doi{10.1109/TCSI.2013.2283994}.

\bibitem[{Tsubone \& Saito(1997)}]{Tsubone1997}
Tsubone, T. \& Saito, T. [1997] \enquote{A {4D} manifold piecewise linear
  hyperchaos generator,}  \emph{Circuits and Systems, 1997. ISCAS '97.,
  Proceedings of 1997 IEEE International Symposium on}, pp. 773--776 vol.2,
  \doi{10.1109/ISCAS.1997.621827}.

\bibitem[{Volos \emph{et~al.}(2017)Volos, Maaita, Vaidyanathan, Pham,
  Stouboulos \& Kyprianidis}]{Volos2017}
Volos, C., Maaita, J.~O., Vaidyanathan, S., Pham, V.~T., Stouboulos, I. \&
  Kyprianidis, I. [2017] \enquote{A novel four-dimensional hyperchaotic
  four-wing system with a saddle-focus equilibrium,} \emph{IEEE Transactions on
  Circuits and Systems II: Express Briefs} \textbf{64},  339--343,
  \doi{10.1109/TCSII.2016.2585680}.

\bibitem[{Wang \& Dong(2015)}]{Wang2015}
Wang, B. \& Dong, X. [2015] \enquote{Secure communication based on a
  hyperchaotic system with disturbances,} \emph{Mathematical Problems in
  Engineering} \textbf{2015}.

\bibitem[{Welch(1967)}]{Welch1967}
Welch, P. [1967] \enquote{The use of fast {F}ourier transform for the
  estimation of power spectra: a method based on time averaging over short,
  modified periodograms,} \emph{IEEE Transactions on Audio and
  Electroacoustics} \textbf{15},  70--73.

\bibitem[{Wolf \emph{et~al.}(1985)Wolf, Swift, Swinney \& Vastano}]{Wolf1985}
Wolf, A., Swift, J.~B., Swinney, H.~L. \& Vastano, J.~A. [1985]
  \enquote{Determining {L}yapunov exponents from a time series,} \emph{Physica
  D: Nonlinear Phenomena} \textbf{16},  285 -- 317,
  \doi{https://doi.org/10.1016/0167-2789(85)90011-9}.

\end{thebibliography}
\end{document}